\documentclass[aps,prf,groupedaddress]{revtex4-2}

\usepackage[utf8]{inputenc}
\usepackage{amsmath}
\usepackage{xcolor}
\usepackage{graphicx}

\begin{document}

%\preprint{}

\title{On near-field coherent structures in circular and fractal orifice jets}


\author{D. Lasagna}
\email{davide.lasagna@soton.ac.uk}
\affiliation{Faculty of Engineering and Physical Sciences, University of Southampton, Southampton, SO17 1BJ, United Kingdom.}
\author{O. R. H. Buxton}
\affiliation{Department of Aeronautics, Imperial College London, London SW7 2AZ, United Kingdom.}
\author{D. Fiscaletti}
\affiliation{
Faculty of Engineering, University of Bristol, Bristol, BS8 1TR, United Kingdom.\\ 
Delft University of Technology, AWEP Dept., Kluyverweg 1, 2629 HS Delft, The Netherlands.  }


\date{\today}

\begin{abstract}
To investigate the influence of the orifice geometry on near-field coherent structures in a jet, Fourier-Proper Orthogonal Decomposition (Fourier-POD) is applied. Velocity and vorticity snapshots obtained from tomographic particle image velocimetry at the downstream distance of two equivalent orifice diameters are analysed. Jets issuing from a circular orifice and from a fractal orifice are examined, where the fractal geometry is obtained from a repeating fractal pattern applied to a base square shape. While in the round jet energy is mostly contained at wavenumber $m=0$, associated to the characteristic Kelvin-Helmholtz vortex rings, in the fractal jet modal structures at the fundamental azimuthal wavenumber $m=4$ capture the largest amount of energy. In addition, energy is scattered across a wider range of wavenumbers than in the round jet. The radial Fourier-POD profiles, however, are nearly insensitive to the orifice geometry, and collapse to a universal distribution when scaled with a characteristics radial length. A similar collapse was recently observed in POD analysis of turbulent structures in pipe flow. However, unlike in pipe flow, the azimuthal-to-radial aspect ratio of the Fourier-POD structures is not constant and varies greatly with the wavenumber. The second part of the study focuses on the relationship between streamwise vorticity and streamwise velocity, to characterise the role of the orifice geometry on the lift-up mechanism recently found to be active in turbulent jets (\textcite{Nogueira2019}). The averaging of the streamwise vorticity conditioned on intense positive fluctuations of streamwise velocity reveals a pair of vorticity structures of opposite sign flanking the conditioning point, inducing a radial flow towards the jet periphery. This pair of structures is observed in both jets, even if the azimuthal extent of this pattern is $30\%$ larger in the jet issuing from the circular orifice. The coupling between streamwise vorticity and velocity motions is also examined using Fourier-POD. The analysis reveals that in the jet with circular orifice lower wavenumber modes, corresponding to structures at larger scales, capture a larger fraction of the vorticity-velocity coupling. This evidences that the orifice geometry directly influences the interaction between velocity and vorticity.
\end{abstract}

%\keywords{}

\maketitle

\section{Introduction}

Coherent structures can be defined as organised fluid elements that capture the overall flow dynamics and are responsible for the transfer of mass, momentum and energy \citep{Hussain1986}. The energetic coherent structures in the near-field of a jet directly affect the production of acoustic noise, the entrainment of quiescent fluid, and the laminar-turbulent transition. For the implications that these aspects have on a range of engineering problems, coherent structures in the near-field of jets have been the subject of several investigations over the last decades. 

In their pioneering work, \textcite{Glauser1987} applied Proper Orthogonal Decomposition (POD) \citep{lumley1967} to the jet mixing layer at three nozzle diameters downstream of the jet exit. They showed that the first POD mode contains $40\%$ of the total turbulent kinetic energy, with an additional $40\%$ from a combination of the next two modes. Later, \textcite{Citriniti2000} expanded from these observations by applying POD to the instantaneous fields of streamwise velocities obtained from an array of 138 hot-wire anemometer probes, at the same downstream location. They observed the existence of azimuthally coherent ``volcano-like'' bursting events, which were short-lived even if containing most of the energy, separated by a ``braid'' region of streamwise counter-rotating vortices. This scenario is consistent with the flow visualisations of \textcite{Liepmann1992} (see figures 7 to 11 of their work). 

The experimental study of \textcite{Jung2004} examined the effects of both downstream position and Reynolds number on the energetic coherent structures of a jet. The total energy content of the modes at azimuthal wavenumber $m=0$ decreases with the distance from the nozzle, while the energy distribution for the first POD mode does not depend on the Reynolds number. The ``volcano-like'' eruptions are dominant from $x/D=2$ to $x/D=4$, but beyond 4 diameters they die off. \textcite{Gamard2004} showed that in the far-field region of the jet the streamwise velocity fluctuations stabilise asymptotically to the azimuthal wavenumber $m=2$. Later, \textcite{Iqbal2007} studied the near-field coherent structures of a jet using the three velocity components from hot-wire signals. From the projection of the first POD mode onto the two-components instantaneous realisations, the local dynamic behaviour of the coherent structures was determined, revealing a helical vortical structure in the range between 4 and 6 diameters. In the studies examined thus far, the jets under analysis can be assumed incompressible. Experimental works with laser doppler anemometry (LDV) and particle image velocimetry (PIV) showed that compressibility effects do not modify the near-field large-scale coherent structures \citep{Taylor2001,Tinney2008}. More recently, the existence of elongated streaky-structures was found using a combination of Spectral POD, Resolvent Analysis, and transient growth analysis (\textcite{Nogueira2019}). These streaks are analogous to those observed by \textcite{Hellstrom2016} in turbulent pipe flows and by \textcite{Hutchins2007} in a boundary layer. \textcite{Nogueira2019} also found that these streaks are characterized by a ratio between the streamwise and azimuthal length scales remaining constant as the azimuthal wavenumber is varied.

Although the coherent structures and their energy content are relatively well-known aspects for a jet with a circular nozzle, different nozzle geometries have scarcely been investigated, and mainly in relation to mixing and entrainment. \textcite{Chrighton1974} was among the first to examine how the nozzle geometry affects the instabilities in the near field of a jet, using linear stability analysis. Later, \textcite{Ho1987} conducted experiments on a jet from an elliptic nozzle at a low aspect ratio. The mass entrainment rate was observed to increase as compared with a round jet, while the mean flow properties in the near field, such as spread and momentum thickness, were found to be different in the planes of the major and minor axis. Later, \textcite{Husain1993} examined an elliptic jet, both numerically and experimentally. The formation of azimuthally-fixed streamwise vortices was observed, which enhance entrainment and mixing when compared with the round jet. The use of tabs in the jet nozzle as a passive means of flow control received also large attention, both in relation to mixing \cite{Foss1999} and entrainment \cite{Zaman1997}. Tabs lead to an increase of the mixing over a wide range of scales, and to a general enhancement of the entrainment rate. \textcite{Gutmark1999} provide a detailed survey on how deviations from circularity affect the near-field coherent structures, which then result in a change of the mean flow properties. 

One of the first attempts aimed at deriving a low-dimensional representation of a jet from a non-circular nozzle is the work by \textcite{Moreno2004}, who applied POD to planar velocity fields of a supersonic rectangular jet, from PIV measurements. Most of the fluctuation energy is contained within the first two modes. Low-order modelling confidently predicted the global flow characteristics, although the effects of the rectangular nozzle were not discussed in relation with the circular nozzle. To identify the individual contributions of the different vortical structures to entrainment, \textcite{ElHassan2010} performed a POD analysis of jets from circular orifice and from  daisy-shaped orifice, and \textcite{ElHassan2011} investigated the near field of a cross-shaped orifice jet using a similar approach. These investigations were conducted with time-resolved stereoscopic PIV and evidenced that the Kelvin-Helmholtz dynamics play a central role in the entrainment of the circular jet. The braid region produces the largest level of entrainment, while the front part of the Kelvin-Helmholtz ring tends to expand the flow and dramatically reduce the entrainment, even in presence of strong streamwise vortex pairs generated by the lobes of the daisy-shaped orifice. 

Lobed nozzles were found to increase turbulent mixing over a wide range of length scales \cite{Hu2001, Hu2002a, Hu2002b}. The involvement of multiple scales in the mixing process is a consequence of the break down of the large-scale streamwise vortices into smaller but not weaker vortices. The enhancement of mixing produced by the lobes is particularly strong in the first diameter, while it becomes almost negligible after the first two diameters. \textcite{Mao2006} and \textcite{Mao2009} studied experimentally the effects of the lobe geometry on the strength of the streamwise vortices in relation to mixing. Lobes having parallel side walls, i.e. rectangular lobes, were found to generate stronger streamwise vortices and hence better mixing performance than triangular and scalloped geometries. 

In the studies discussed up until here, nozzle geometries different from circular were meant to enhance mixing or entrainment. However, the reduction of the aeroacoustic emissions represented the main motivation for a number works on jet from non-circular orifices. Among these, the experimental work of \textcite{Tam2000} examined the far-field noise from elliptic, rectangular, lobed, and tabbed jets. While the noise radiated from elliptic and rectangular jets is the same as that from the  circular jet, a significant suppression of the large-scale noise is obtained from the lobed jet. Tabs also impact on the noise field primarily by shifting the spectral peak to a higher frequency. 
With the aim of investigating the screeching in a rectangular jet, \textcite{Alkislar2003} performed stereoscopic PIV. Coherent vortical structures at increasing strength in the shear layer are associated with screeches of  progressively stronger intensity. Later, stereoscopic PIV and microphones were used to investigate the aeroacoustics effects of chevron and microjets \textcite{Alkislar2007}. 
The low frequency noise is attenuated using both the flow control techniques, while the high frequency noise tends to increase. The attenuation mechanism of both chevron and microjets was associated with the formation of streamwise vortices disrupting the generation of azimuthally coherent large-scale structures. 

The time evolution of the near-field structures in a jet both from a circular nozzle and from a chevron nozzle has been studied using time-resolved tomographic PIV (\textcite{Violato2011}). In that work, the intense vortical structures were identified with criteria based on the analysis of the velocity gradient tensor. 
% thresholds of vorticity and of $\lambda_{2}$ \citep{Jeong1995}. 
From Powell's acoustic analogy, the pairing of the vortex rings represents the main source of noise in the circular jet. On the other hand, vortex rings are nearly absent in the chevron jet, while counter-rotating streamwise vortices develop from the chevron notches. The decay of these streamwise vortices leading to the formation of C-shaped structures is regarded as the main mechanism for noise production in the chevron jet. The same dataset was also used to examine with POD the spatial organization of the structures at the jet core breakdown \citep{Violato2013}.

Recently, \textcite{Sinha2016} applied linear stability theory to derive the instability modes and their downstream evolution both in circular and in  chevron jets. The serrated nozzle reduces the growth rates of the most unstable eigenmodes of the jet, although their phase speeds are approximately similar. Coherent structures in the near field of a circular jet were investigated by \textcite{Lesshafft2019}, who applied spectral POD on velocity fields from time-resolved stereoscopic PIV \cite{Jaunet2017}. In the range of Strouhal number around 0.4, the leading modes of spectral POD obtained from experimental data and from resolvent-based modeling show a very good agreement. 
\textcite{Rigas2019} compared the spectral POD modes from a chevron jet and from a circular jet, both obtained from large-eddy simulations. The analysis identified structures that take the form of elongated streamwise streaks, analogous to those observed by \textcite{Nogueira2019}. These streaks have been associated with the non-modal lift-up mechanism in wall-bounded flows. In the circular jet, the energetic streaks appear at the azimuthal wavenumber $m=1$, while in the chevron jet the streaks form in consequence of the nozzle geometry, and they inherit the periodicity of the nozzle geometry itself. The relative importance of the lift-up, the Kelvin-Helmholtz, and the Orr mechanisms was recently investigated by \textcite{Pickering2020} from Large-Eddy Simulations of a round jet. The work points at  the lift-up mechanism as an important linear amplifier of disturbances in turbulent jets, thus confirming and expanding the findings by \textcite{Nogueira2019}. In jets at Mach numbers 0.4, 0.9, and 1.5, the lift-up mechanism was found to be responsible for the generation the streamwise streaks at low-frequency and non-zero azimuthal wavenumbers.  

From the analysis of the past literature, it emerges that the non-circular orifice geometries that were examined were all characterised by only one spatial length scale. It is however not completely understood how an orifice geometry constructed using multiple length scales affects the near-field coherent structures in a jet. Furthermore, in the majority of these investigations the focus was on mixing and on entrainment. To the authors' best knowledge, an analysis of the lift-up mechanism in jets issuing from a non-circular orifice has never been performed. The novel contribution of this work is that we examine and compare near-field coherent structures in jets with two different orifice geometries, i.e.~a round orifice and a fractal orifice, using Fourier Proper Orthogonal Decomposition as the primary tool for investigation. Although a fractal orifice may not appear to be of direct interest to industry, this geometry embedding a wide range of length scales can be used to efficiently assess the sensitivity of the near-field coherent structures to the flow initial conditions. The velocity fields under analysis are obtained from planar and tomographic PIV at high spatial resolution used in previous works \citep{Breda2018b, Breda2019}. Measurements at two equivalent nozzle diameters from the exit are considered. In the first part of the study, we quantify the role of the orifice geometry on the energy distribution across the hierarchy of modal structures from the three velocity components. The properties of self-similarity of these modal structures are examined for the two orifice jets, both along the radial and the azimuthal directions. Analogies and dissimilarities between POD modes observed in these jets and those recently found in pipe flows are discussed. In the second part of the study, the structures of streamwise vorticity leading to the near-field velocity streaks (\textcite{Nogueira2019}) are investigated. The focus is on the effects of the orifice geometry on the spatial characteristics of these vorticity structures and on the lift-up mechanism of streaks formation. 

\section{Experimental datasets}

The jet flow was generated by an open jet facility at Imperial College London, described elsewhere \cite{Breda2018b, Breda2019} (see figure 2 of Ref.~\cite{Breda2018b} for details on the nozzle  geometry). In order to prevent biasing the particle images with unseeded, quiescent air being entrained into the jet, a seeded, mild co-flow of air was applied. The exit flow was found to have a sharp 'top-hat' mean velocity profile and a turbulence intensity $< 1 \%$. 
\begin{figure}
\centering
    \includegraphics[width=0.62\textwidth]{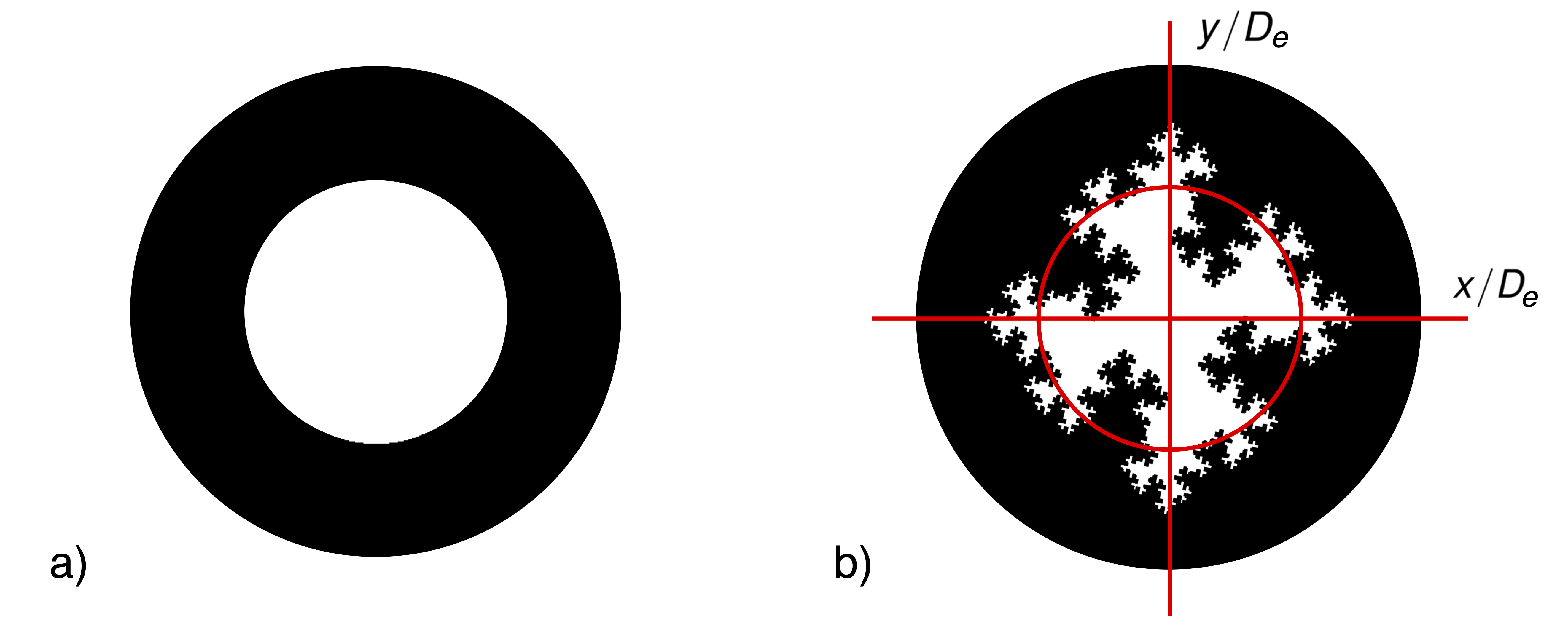}
    \caption{The orifice geometries considered in this study, (a) circular and (b) fractal. The red circle in panel (b) traces the outline of the circular orifice in panel (a).}
    \label{fig:nozzle}
\end{figure} 
A round orifice and a fractal orifice were used as illustrated in Fig.~\ref{fig:nozzle}, where the fractal geometry is obtained from a repeating fractal pattern applied to a base square shape and has a fractal dimension of 1.5 and 3 iterations. This pattern substantially increases both the wetted perimeter and the number of corners. A similar geometry has been previously adopted for the perimeter of axisymmetric wake-generating plates where a break-up of the coherence and a reduction of the shedding energy were observed by \textcite{Nedic2015}. The two orifices have same open area of $D_e^{2}$, where $D_e = 15.78$ mm is the equivalent diameter as defined in \textcite{Breda2018a}.
% This was obtained from two different orifices, round and fractal, of thickness of $0.1$ mm and of identical open area of $\pi D^{2}_\textup{e}/4$, where $D_\textup{e} = 15.78$ mm is the equivalent diameter. 
% For a given open area of an orifice, a repeating fractal pattern is one that maximises both the wetted perimeter and the number of corners.
% The geometry of the fractal nozzle under analysis is characterised by a fractal dimension of 1.5 and 3 iterations (see figure \ref{fig:nozzle}(b)).
The two jets have the same exit velocity $U_j = 9.93$ $m s^{-1}$, meaning that the Reynolds number $Re_{D_e} = {U_j D_e}/{\nu}$ is the same and is equal to $10^{4}$.

A Cartesian coordinate system is introduced, centred on the geometric centre of the orifices, with the $z$-axis oriented along the streamwise flow direction and where the $x$- and $y$-axes are aligned with the diagonals of the fundamental square pattern for the fractal orifice geometry, as illustrated in Fig.~\ref{fig:nozzle}(b). In polar coordinates, radial and azimuthal directions are indicated by $r$ and $\theta$, respectively, with the latter originating on the $x$-axis. Velocity fluctuations along the three Cartesian directions are denoted as $u_x, u_y$ and $u_z$, while time averaged quantities are denoted by an overbar. In polar coordinates, radial and azimuthal velocity fluctuations are denoted as $u_r$ and $u_\theta$, respectively.

Three datasets from particle image velocimetry (PIV) are analysed in this work, \textit{i.)} a dataset from planar PIV at higher resolution (HPIV), \textit{ii.)} a dataset from planar PIV at lower  resolution but at larger field of view (LPIV), and \textit{iii.)} a dataset from tomographic PIV (TPIV).  To obtain the HPIV dataset, a final interrogation area of $12 \times 12$ pixels with a $50\%$ overlap was used in the processing. This led to a vector spacing of $0.013D_{e}$, with a spatial resolution of less than $3.7\eta$, where $\eta$ is the Kolmogorov length scale. The LPIV dataset enabled the study of the flow in the range between $0$ and $23D_{e}$, as shown in figure 2.5 of \textcite{Bredathesis2018}. The fields of view were $9D_{e} \times 6.7D_{e}$. An initial interrogation area of $64 \times 64$ pixels was used to process the images hierarchically, with a final window size of $16 \times 16$ pixels. A $50\%$  overlap between adjacent interrogation areas was used leading to a final vector spacing of $0.03D_{e}$. Additional details on the two datasets of planar PIV can be found in \textcite{Breda2018b} and in \textcite{Bredathesis2018}. 
The three-dimensional velocity fields under analysis were obtained with tomographic particle image velocimetry (TPIV) at two orifice diameters downstream of the orifice. The interrogation volume was of $48 \times 48 \times 48$ voxels, with a $75 \%$ window overlapping. The spatial resolution in the worst case was of $11 \eta$. The processing of the PIV images from the three datasets was done with the software DaVis by LaVision. Additional information on the TPIV measurements can be found in \textcite{Breda2019}, in section 2.3.

\section{Mean flow properties}
\label{sec:meanflow}
\begin{figure}
\centering
    \includegraphics[width=1\textwidth]{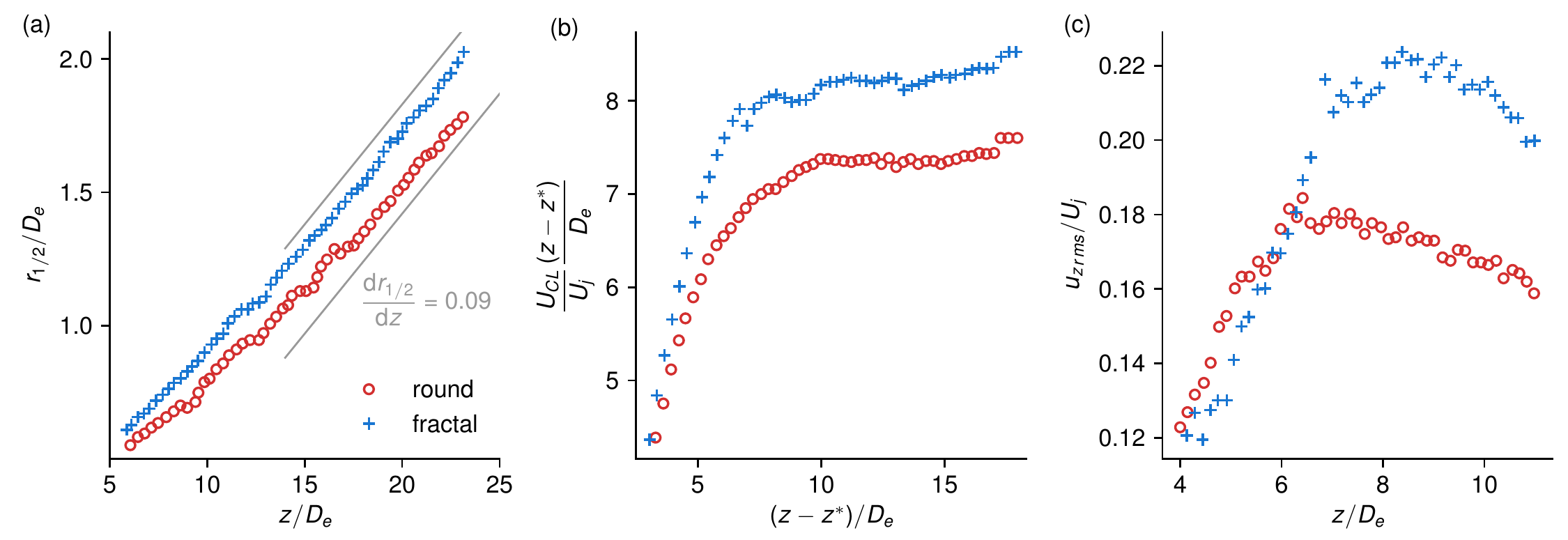}
    \caption{Mean flow properties for both the round jet (red circles) and the fractal jet (cyan crosses). (a) Spreading rate, (b) streamwise velocity decay, (c) turbulence intensity. }
    \label{fig:mean_streamwise}
\end{figure}

In this section, mean flow properties of the two jets are illustrated and discussed. The spreading rate of the two jets is first examined. From Fig.~\ref{fig:mean_streamwise}(a), the two jets exhibit analogous spreading rates. Specifically, the spreading rate is around $\mathrm{d} r_{1/2}/\mathrm{d}z=0.09$, where $r_{1/2}$ is the jet half-width. This result is similar to previous measurements on a jet issuing from circular nozzle (see \textcite{Panchapakesan1993} and \textcite{Hussein1994}).  However, Fig.~\ref{fig:mean_streamwise}(a) shows that the fractal jet exhibits a larger half-width than the round jet.

In Fig.~\ref{fig:mean_streamwise}(b), the streamwise velocity decay is shown, where $U_{CL}$ is the jet centerline velocity and $z^*$ is the jet virtual origin \cite{Pope2000}. Beyond a distance from the orifice of ten equivalent diameters, the streamwise velocity decay becomes approximately constant. The value of the velocity-decay constant is not the same in the two jets, which is an important difference in the mean flow properties. Specifically, a larger decay is observed in the fractal jet. This is a direct consequence of the shorter potential core of the round jet when compared to the fractal jet, as shown by \textcite{Breda2018b}. 
%The slower downstream development of the fractal jet is discussed in further details later in the paper. 
Additional evidence that the potential core of the round jet is shorter than that of the fractal jet comes from examining the streamwise turbulence intensity $u_{z_{rms}}/U_j$ at the centerline, shown in Fig.~\ref{fig:mean_streamwise}(c). In fact, the turbulence intensity in the centerline of the round jet presents a peak at $z/D_{e} \approx 6$ before starting to decay, whereas for the fractal jet velocity fluctuations saturate at a later distance from the orifice, at around $z/D_{e} \approx 9$. 

\begin{figure}
\centering
    \includegraphics[width=1\textwidth]{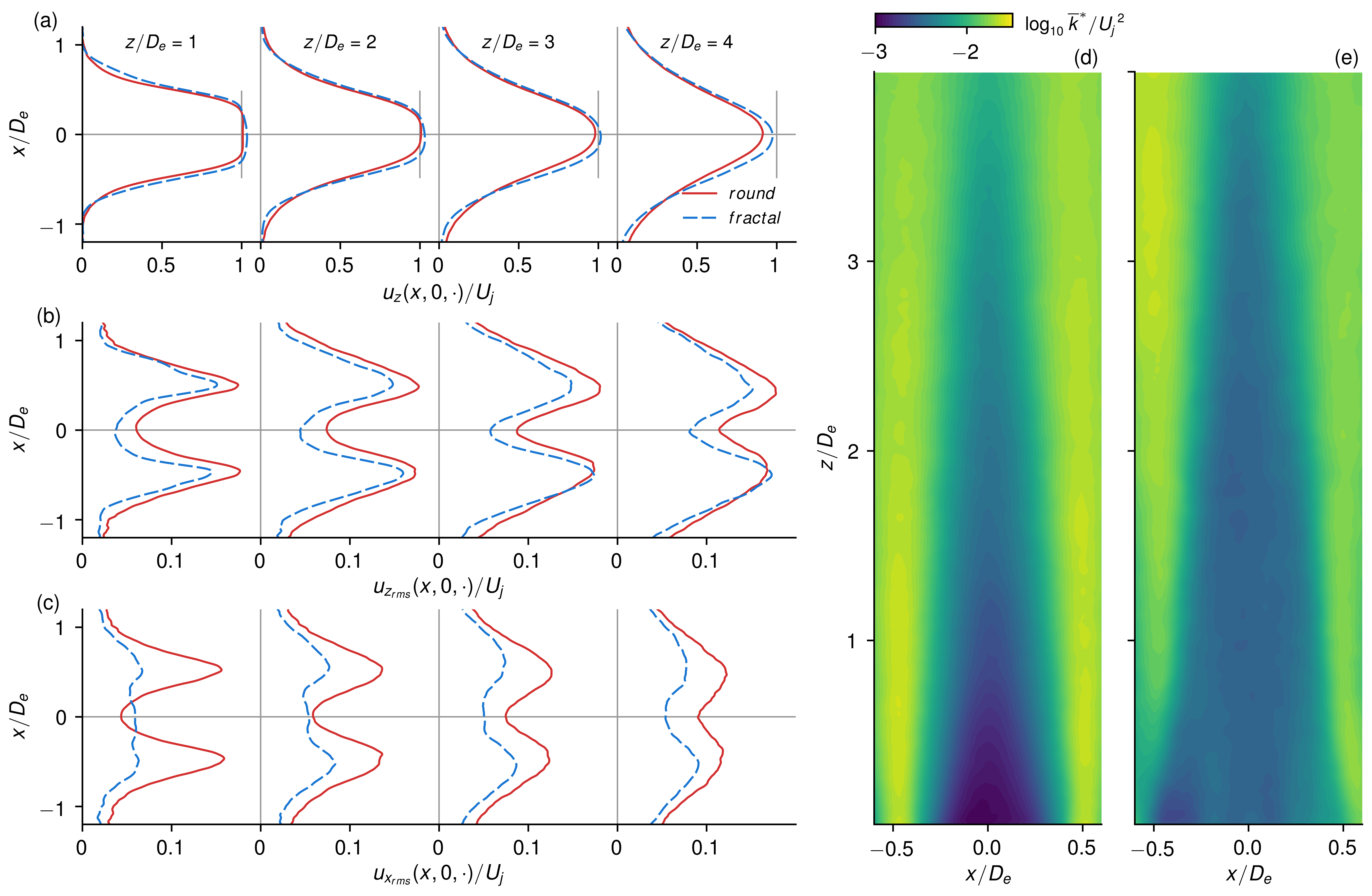}
    \caption{In the first column, (a) mean streamwise velocity profiles at four different streamwise distances from the orifice, for the fractal (red continuous line) and round (cyan dashed line) orifices; (b,c) profiles of the streamwise and transversal rms velocity, respectively, for the two orifices at the same streamwise locations. In the second column, planar turbulent kinetic energy $\overline{k}^* \equiv \overline{(u_z^2 + u_x^2)}/2$ normalised by the local streamwise velocity, for (d) the round jet and (e) the fractal jet. }
    \label{fig:planar-profiles}
\end{figure}

Radial profiles of the streamwise velocity of the round jet (continuous red line) and of the fractal jet (dashed cyan line) are presented in Fig.~\ref{fig:planar-profiles}(a), at the same four different downstream locations from the orifice, i.e. at $z/D_{e}=1$, $2$, $3$, and $4$. Over this range, the velocity profiles undergo a transition from a top-hat shape to nearly Gaussian, and exhibit a radial spread. 
It can also be observed that at $z/D_{e}=1$ and $2$ both velocity profiles present a mild overshoot, as they reach values of $u_z(x,0,\cdot)/U_{j}>1$. This can be explained as a \textit{vena contracta} effect in proximity of the orifices. At $z/D_{e}=4$, the centerline velocity of the fractal jet is larger than that of the round jet, suggesting that the jet spreads radially more pronouncedly in the round jet, despite the strong mixing action induced by the irregular orifice geometry. This observation is consistent with \textcite{Breda2018b}. 

% NOTA: - the comparison between circular and fractal jets is likely biased by an earlier transition to turbulence in the latter case, as may be observed in the visualisations in Breda and Buston PoF 2018. The round jet is potentially in a region with reminiscent transition effects, hence the dominance of m=0, which is not observed by turbulent jets, at least for higher Re. 

The radial profiles of the streamwise velocity fluctuation rms are given in Fig.~\ref{fig:planar-profiles}(b), at the four different downstream locations from the orifice. In general, more intense fluctuations in the near field region are observed in the round jet. At $z/D_{e}=1$, the profiles exhibit two local maxima in proximity of the shear layer and a local minimum at the centerline. At growing downstream locations, the local maxima tend to become weaker, whereas the local minimum strengthens. The increase of the turbulence levels at the centerline is associated with the thickening of the shear layer at increasing downstream locations. Fig.~\ref{fig:planar-profiles}(c) shows profiles of the rms transversal velocity, i.e. of the rms of the velocity component along the $x$-axis (see Fig.~\ref{fig:nozzle}). These profiles evidence a much stronger discrepancy between the round jet and the fractal jet than the streamwise velocity rms profiles observed in Fig.~\ref{fig:planar-profiles}(b). In fact, the transversal velocity fluctuations in the fractal jet are significantly less intense, especially at $z/D_e = 1$. Such an attenuation is consistent with the delay of the streamwise velocity decay found in the fractal jet, and with a potential core extending over a longer streamwise distance. It is anticipated here, and fully detailed in the Fourier-POD analysis of the velocity fluctuations at $z/D_e=2$ reported in section \ref{sec:podcomponents}, that this reduction is associated to the suppression of the strong Kelvin-Helmholtz instabilities featuring in the near field of the round jet, which strongly couple the streamwise and transversal velocity components.
% Therefore, from the analysis of the near-field statistics, we can infer that the most evident effect of the fractal orifice in comparison to the round orifice is to suppress the turbulence fluctuations in the direction orthogonal to the centerline. 
The planar turbulent kinetic energy in the near field of the fractal and round jets is reported in Figs.~\ref{fig:planar-profiles}(d) and  \ref{fig:planar-profiles}(e), respectively. These maps confirm that the potential core of the fractal jet presents a longer downstream development. This finding is consistent with the downstream development of the centerline turbulence intensity in Fig.~\ref{fig:mean_streamwise}(c), and with the estimates of the number of eddies overturned in time for a given downstream distance estimated in  \textcite{Breda2018b} (their figure 7). 

From the discussion of the results presented up until here, the fractal orifice leads to the following modifications of the near-field jet structure when compared to the circular orifice: 1) an increase of the streamwise extent of the potential core; 2) a reduction of the decay rate of the streamwise velocity; 3) a strong attenuation of the transversal velocity rms and a mild attenuation of the streamwise velocity rms. These experimental findings are consistent with recent results from linear stability analysis presented in \textcite{Lyu2019} and in \textcite{Lajus2019} on jets issuing from non-circular (lobed) nozzles. These two studies concluded that the temporal growth of near-field instabilities decreases for increasing number of lobes. Additionally, \textcite{Lyu2019} observed that also a larger penetration ratio of the lobes of the jet nozzle produces a decrease in the temporal growth rate. 

\begin{figure}
    \includegraphics[width=0.9\textwidth]{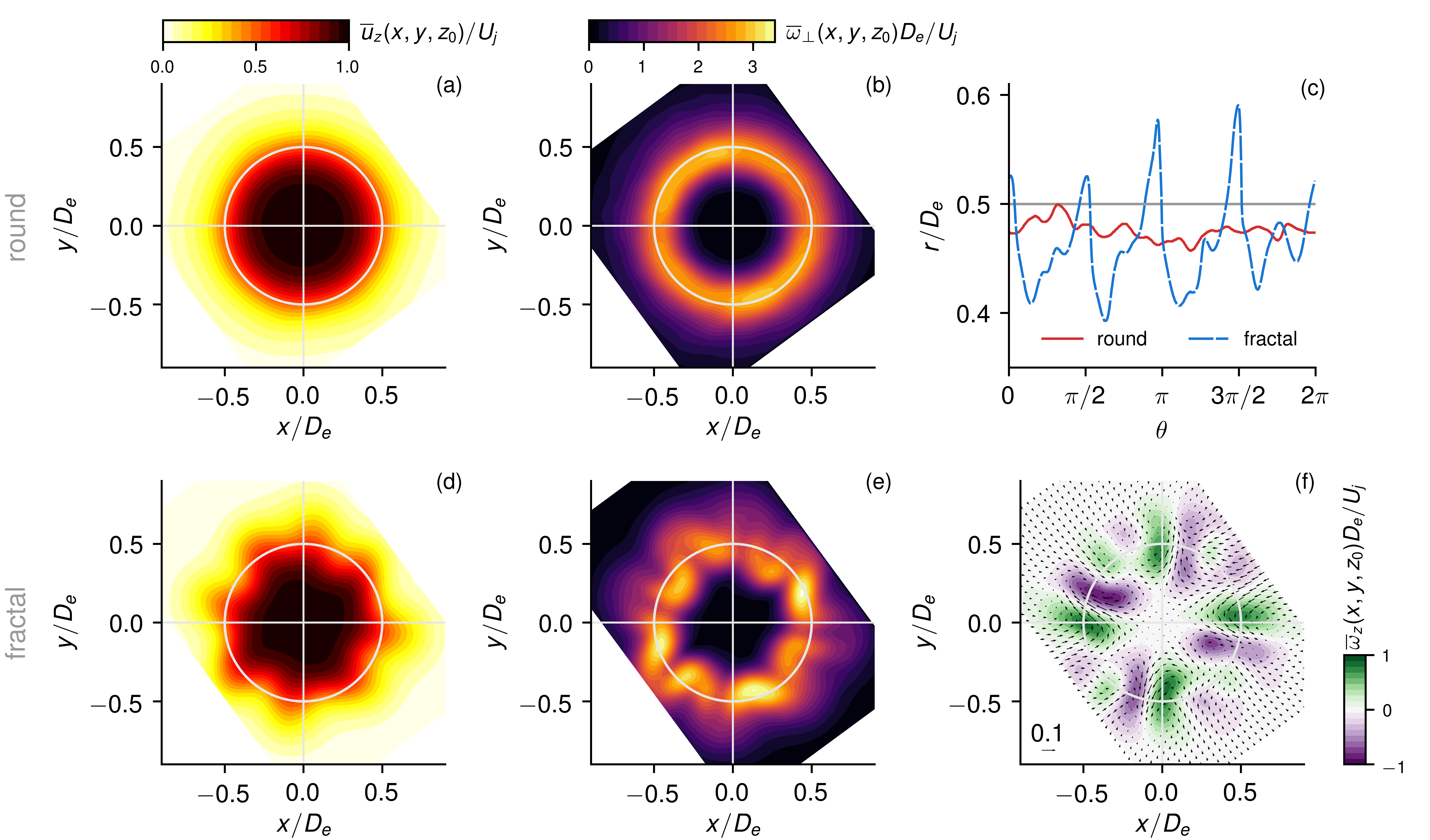}
    \caption{(a,c) Maps of the mean streamwise velocity for (a) the round jet and (d) the fractal jet; (b,e) maps of the mean planar vorticity for (b) the round jet and (e) the fractal jet; (c) radial location of the maximum of $\bar{\omega}_{\bot}$ as a function of the azimuth $\theta$ for both jets; (f) mean velocity components $\bar{u}_{x}$ and $\bar{u}_{y}$ on the $xy$ plane and map of the mean streamwise vorticity $\bar{\omega}_{z}$ for the fractal jet. }
    \label{fig:average-flow-vort}
\end{figure}

\begin{figure}
    \includegraphics[width=0.9\textwidth]{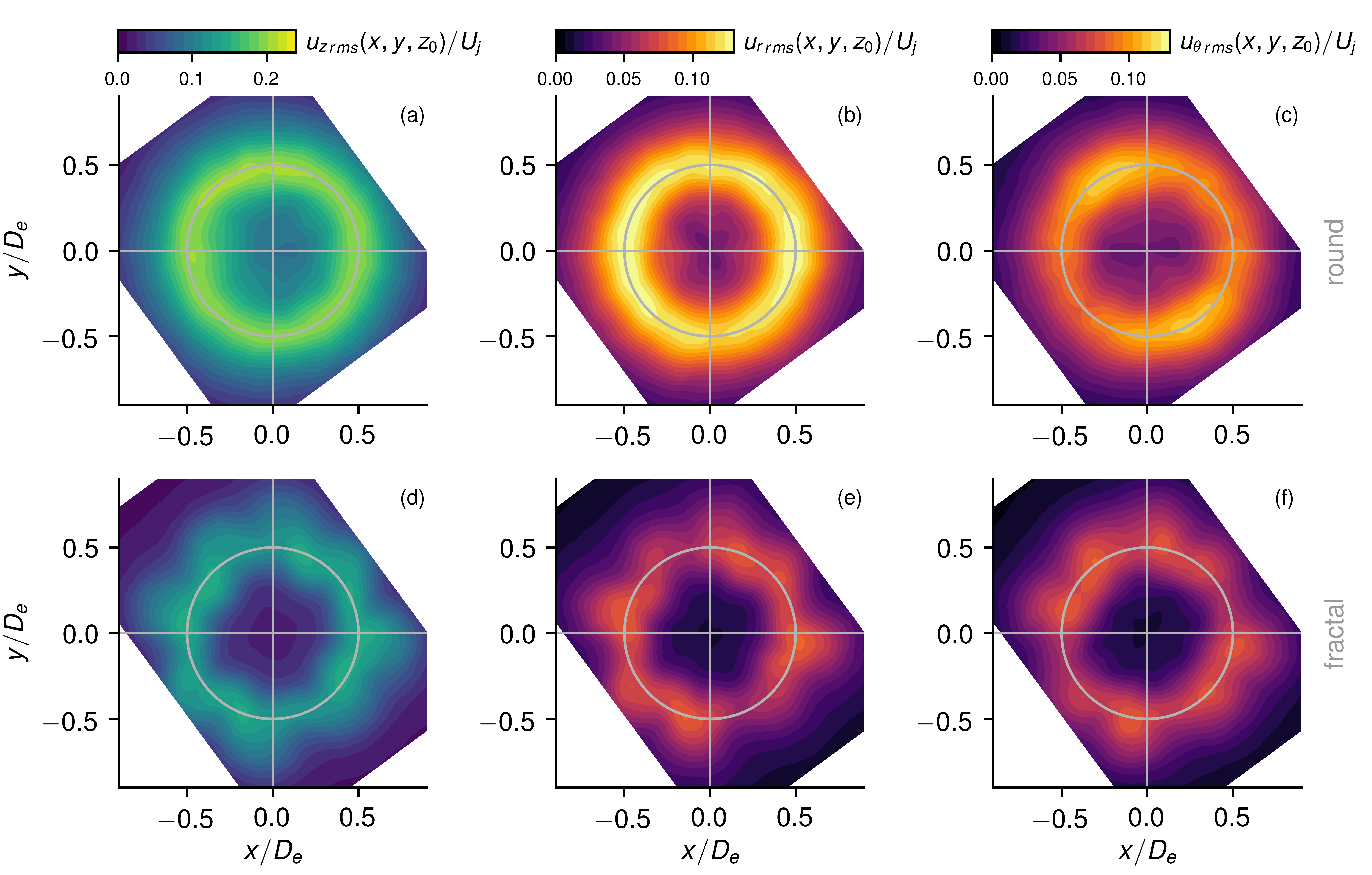}
    \caption{Maps of the root mean square (rms) of the (a, d) streamwise, (b, e) radial, and (c)(f) azimuthal velocity components for (a, b,c ) the round jet (top row) and (d, e, f) the fractal jet (bottom row). }
    \label{fig:average-flow}
\end{figure}

We now focus on the TPIV dataset and present maps of the normalised mean streamwise velocity $\bar{u}_z(x, y, z_\textup{0})$ at a downstream distance from the orifice of $z_\textup{0}/D_e = 2$. These are shown in Figs. \ref{fig:average-flow-vort}(a, d) for the round and fractal orifices, respectively. The fractal geometry produces a significant spatial modulation of the mean velocity distribution in the near field, breaking the azimuthal symmetry of the vortex sheet that is otherwise observed for the round jet. This effect is even more evident when looking at the norm of the mean planar vorticity $\bar{\omega}_{\bot}=\sqrt{\bar{\omega}_{r}^2 + \bar{\omega}_{\theta}^2}$, in Figs. \ref{fig:average-flow-vort}(b,e), with $\bar{\omega}_{r}$ and $\bar{\omega}_{\theta}$, the mean radial and azimuthal vorticity components. The location of the maximum of the planar vorticity along the radial direction, Fig.~\ref{fig:average-flow-vort}(c), reveals that important differences exist between the two orifices. Although in the round jet the mean planar vorticity is maximum in the proximity of the orifice lip line, at a nearly constant radial position, in the fractal jet the radial location of the planar vorticity peak varies with the azimuth. In particular, the radial location varies between $r/D_e=0.4$ and $r/D_e=0.6$. Associated to the corrugation of the mean vortex sheet, a modulation of the mean streamwise vorticity $\bar{\omega}_z$ is also observed, as shown in Fig.~\ref{fig:average-flow-vort}(f), in the form of pairs of adjacent positive-negative patches, elongated along the radial direction.  Each of these pairs is associated with a distinct lobed structure from the map of the mean streamwise velocity shown in Fig.~\ref{fig:average-flow-vort}(d), resulting from secondary flows in the plane of the figure. In total, eight pairs of positive-negative patches are formed, four of which characterised by a stronger intensity (see \textcite{Breda2018a} for a deeper discussion). It is remarkable that the mean streamwise vorticity within these patches is as large as one third of the mean planar vorticity in Fig.~\ref{fig:average-flow-vort}(e). In wall bounded turbulent flows, spanwise inhomogeneities of surface attributes were found to produce streamwise-elongated vorticity structures (see \textcite{Pujals2010}, \textcite{Kevin2017}, \textcite{Vanderwel2019} among others), resulting from the anisotropy of turbulent stresses \cite{perkins_1970, anderson_barros_christensen_awasthi_2015}. In laminar flows, streamwise-elongated vortices were found to delay transition to turbulence \cite{Fransson2005, Fransson2006} in boundary layers and to reduce growth rates of Kelvin-Helmholtz-type instabilities in free shear flows (see \textcite{Lajus2015}, \textcite{Boujo2015}, \textcite{Marant2018} among others). In the present case, the near-field flow in the fractal jet is highly turbulent and strongly influenced by the irregular geometry of the nozzle. However, as it will be extensively quantified in the next sections using a Proper Orthogonal Decomposition, a significant reduction of the strength of azimuthally coherent velocity fluctuations in the fractal jet is observed \cite{Breda2018a}, which is akin to mechanisms observed in laminar flows.
% This is consistent with the characteristics of the mean flow properties of the two jets discussed in section \ref{sec:meanflow}, and with their turbulence statistics discussed in section \ref{sec:meanflow}. 
% A deeper analysis on the coupling between streamwise vorticity and  streamwise velocity is provided in section \ref{sec:coupling}, where the length scales involved in this coupling are quantified. 

The strong shear between the jet flow and the quiescent ambient fluid is responsible for an intense production of turbulent kinetic energy. To examine how the orifice geometry affects the spatial distribution of the different terms constituting the turbulent kinetic energy, we show in Fig.~\ref{fig:average-flow} maps of the rms value of the (a, d) streamwise, (b, e) radial, and (c, f) azimuthal velocity components, respectively, for the (top row) round jet and for the (bottom row) fractal jet. Among the three components, the streamwise velocity is dominant. For this component, the rms value is significantly lower in the fractal jet, consistent with our observations from the planar PIV datasets presented in Fig.~\ref{fig:planar-profiles}(b). As mentioned above, it is argued that this effect can be attributed to the strong deviation from the azimuthal symmetry of the fractal orifice, which reduces the growth of Kelvin-Helmholtz instabilities and the near-field jet development. A lower intensity of Kelvin-Helmholtz-induced fluctuations also explains the significantly lower levels of radial velocity rms in the fractal jet, since azimuthally coherent vortex rings also induce strong radial motions \cite{Liepmann1992}. This can be observed in Figs.~\ref{fig:average-flow}(b) and \ref{fig:average-flow}(e). We note that, although not shown here, the correlation coefficient between the radial and streamwise components in the region of intense shear is comparable in the two jets. The coefficient, approximately 0.6, is positive, since a positive (outward) radial motion induces a positive streamwise fluctuation by exploiting the high mean velocity gradients in the near field. On the other hand, the rms of the azimuthal velocity appears to be milder than that of the other two components, and of comparable strength in both jets, as shown in Figs.~\ref{fig:average-flow}(c) and (f). 

\section{Proper orthogonal decomposition: method of analysis}
\label{pod-method}
We use Proper Orthogonal Decomposition to identify near-field coherent structures in an objective manner and elucidate the role of the orifice geometry on the structure of turbulence.
% by finding an orthonormal set of elementary basis functions ${\boldsymbol \phi}_i(r, \theta, z) \equiv (\phi^i_r, \phi^i_\theta, \phi^i_z)$ that is optimal in the sense that the average projection of the turbulent velocity field onto this set is maximum. 
We follow the standard approach for problems defined in cylindrical coordinates \citep{GAMARD:2002eh, Hellstrom2016}, and consider velocity fluctuation vector fields $\mathbf{u}(r, \theta, z, t)$ with radial, azimuthal and streamwise velocity components restricted to a radial-azimuthal plane located at $z_0/D = 2$. For completeness, we provide in what follows a brief description of the methodology, restricted to the scalar implementation where the analysis is performed on each velocity component independently \cite{Tinney2008}. 

A classical variational technique \citep{lumley1967} can be used to derive a complete, orthonormal set of modal structures $\{ \phi^i_j(r, \theta) \}_{i=1}^\infty$, ordered by the modal kinetic energies $\{\lambda^i_j \}_{i=1}^\infty$, from the solution of the integral eigenvalue problem
\begin{equation}
\int_0^{2\pi} \int_0^\infty R_j(r, r^\prime, \theta, \theta^\prime) \phi^i_j(r^\prime, \theta^\prime)r^\prime \mathrm{d}r^\prime\mathrm{d}\theta^\prime = \lambda^i_j\phi^i_j(r, \theta),
\end{equation}
where $R_j(r, r^\prime, \theta, \theta^\prime)$ is the two-point correlation tensor defined by
\begin{equation}\label{eq:two-point-corr-tensor}
	R_{j}(r, r^\prime, \theta, \theta^\prime) = \mathrm{E}\left[ u_j(r, \theta, t) u_j(r^\prime, \theta^\prime, t)\right],
\end{equation}
where $\mathrm{E}[\cdot]$ is the expectation operator, and $j$ identifies the radial ($r$), azimuthal ($\theta$) or streamwise ($z$) velocity component. For space-only POD, as in the present case, the expectation operator is the arithmetic average over the available velocity snapshots. 

For the round jet, the equations of motions and boundary conditions are equivariant under the continuous group of rotations $\mathcal{R}^\beta : \mathbf{u}(r, \theta) \mapsto \mathbf{u}(r, \theta + \beta)$. Turbulence statistics are then homogeneous along the azimuthal coordinate $\theta$ and the two-point correlation tensor $R$ only depends on the azimuthal separation, i.e. $R_{j}(r, r^\prime, \theta, \theta^\prime) = R_{j}(r, r^\prime, \theta - \theta^\prime)$. It is well known that, in this particular case, the POD modes have the azimuthal structure of Fourier modes. This property can be exploited to first Fourier transform the velocity fluctuation snapshots along the azimuthal direction
% , as follows:
% \begin{equation}
    % \mathbf{u}(r, \theta, t) = \sum_{m=-\infty}^{\infty}\mathbf{u}(r, t, m)\mathrm{e}^{i m \theta},
% \end{equation}
% where $i = \sqrt{-1}$, 
and then apply the POD to the complex-valued transformed fields by solving the eigenvalue problem
\begin{equation}\label{eq:integral-equation-fourier}
\int_0^\infty R_j(r, r^\prime, m) \phi_j^i(r^\prime, m)r^\prime \mathrm{d}r^\prime = \lambda^i_j(m)\phi_j^i(r, m),
\end{equation}
for a set of azimuthal wavenumbers $m$, where $R_{j}(r, r^\prime, m) = \mathrm{E}[u_j(r, m, t) u^\dag_j(r, m, t)]$, with $(\dag)$ denoting complex conjugation. To lift the asymmetry of the kernel in the integral equation (\ref{eq:integral-equation-fourier}) introduced by the term $r^\prime$ arising from the energy-based inner product of velocity fields defined over a cylindrical coordinate system, we follow the established approach used in other works on round jets or pipe flow \citep{Iqbal2007, Hellstrom2016}. In what follows, we refer to this approach as Fourier-POD analysis. In practice, we proceeded by 1) interpolating the PIV velocity fluctuation fields onto a polar grid originating at the jet centre (identified from the mean field), 2) Fourier transforming the data along the streamwise direction, 3) constructing the two-point correlation tensor $R_j(r, r^\prime, m)$ and 4) solving the discrete equivalent of the eigenvalue problem of equation (\ref{eq:integral-equation-fourier}).
% is lifted by defining the modified tensor 
% \begin{equation}
% 	\tilde{R}_{ij}(r, r^\prime; m) = \mathbb{E}\left[ \sqrt{r}\hat{u}_i(r, t; m) \hat{u}_j^*(r^\prime, t; m)\sqrt{r^\prime}\right]
% \end{equation}
% where $(\cdot)^*$ denotes conjugation, and then solving 
% \begin{equation}\label{eq:final-pod-problem}
% \int_0^\infty \hat{\mathbf{M}}(r, r^\prime, m) \hat{{\boldsymbol \phi}}^i(r^\prime; m) \mathrm{d}r^\prime = \lambda^n(m) \hat{{\boldsymbol \phi}}^i(r; m),
% \end{equation}where $\lambda^n(m)$ is the turbulent kinetic energy captured by mode $\hat{\boldsymbol{\phi}}^i(m)$. 

% We apply a  Fast Fourier Transform for real data, which only return (approximately) half of the spectrum of Fourier coefficients, for non-negative wave numbers. Hence, to account for the energy at negative wave numbers, the eigenvalues obtained from the problem in equation  \ref{eq:final-pod-problem} for $m>0$ are multiplied by a factor two.

For the jet issuing from the fractal orifice, the equations and boundary conditions are equivariant under a cyclic group of order four generated by the symmetry $\mathcal{T} = \mathcal{R}^{\pi/2}$, rotating velocity fields around the $z$ axis by $\pi/2$. The important consequence is that velocity statistics are still periodic, but are not homogeneous in the azimuthal direction. The two-point correlation tensor (\ref{eq:two-point-corr-tensor}) does not depend only on the azimuthal separation $\theta - \theta^\prime$ and the POD modes do not necessarily have a simple harmonic azimuthal structure. Hence, while the Fourier-POD analysis is optimal for the velocity dataset of the round jet, it inevitably becomes sub-optimal for the fractal orifice dataset. However, using the same analysis for the two geometries has the advantage of enabling a direct comparison of a) the azimuthal wavenumber distribution of the kinetic energy and of b) the radial profiles of the modal structures. To quantify the degree of sub-optimality of the Fourier-POD approach for the jet issuing from the fractal orifice, we also compute and present optimal structures for the fractal orifice geometry by applying the snapshot POD approach \citep{sirovich} to a larger dataset constructed by applying the rotations $\mathcal{T}, \mathcal{T}^2$ and $\mathcal{T}^3$ on each fluctuation velocity field, thus increasing the number of snapshots used in the analysis by a factor of four.

\section{Proper orthogonal decomposition of the three velocity components}
\label{sec:podcomponents}

\begin{figure}
\centering
    \includegraphics[width=0.95\textwidth]{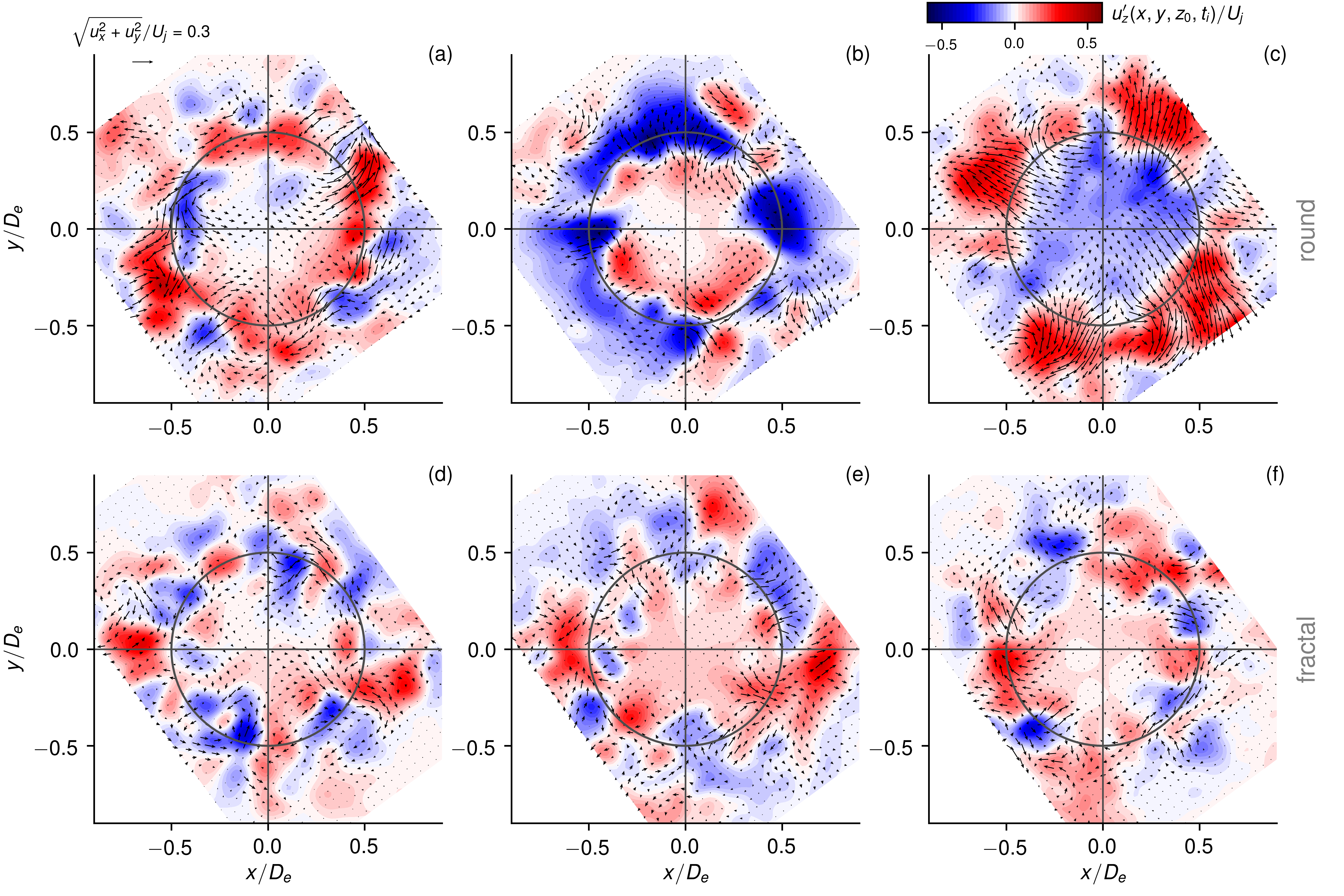}
    \caption{Snapshots of the instantaneous vector fields containing the velocity components $u_{x}$ and $u_{y}$ on the $x-y$ plane and maps of the streamwise velocity fluctuations at $z_0/D_{e} = 2$, for the (a,b,c) round jet and for the (d,e,f) fractal jet. Note that the velocity vector fields of the velocity components $u_{x}$ and $u_{y}$ overlap to the color maps of $u^{\prime}_z$, and they share the same Cartesian axes.}
    \label{fig:snapshots}
\end{figure}

From the analysis of the maps of mean velocity and vorticity examined in previous sections, it was argued that the fractal orifice breaks the azimuthal coherence of the vortex rings that can typically be found in jets issuing from a circular orifice. Therefore, the instantaneous velocity fields of the fractal jet are expected to be populated by structures that are much smaller in size compared with those in the round jet. This should be more evident in regions  characterised by intense shear, therefore in proximity to the lip line. In Fig.~\ref{fig:snapshots}, instantaneous snapshots of the streamwise velocity fluctuations are presented, for the (a, b, c) round jet and for the (d, e, f) fractal jet, to illustrate this behaviour. Velocity vectors in the $x-y$ plane are also shown to illustrate the nature of the cross-plane flow. Footprints of the vortex rings associated with the Kelvin-Helmholtz instabilities passing through the observation plane can be observed in the three snapshots obtained from the round jet. The streamwise velocity fluctuations are organised in large coherent structures developing along the azimuthal direction, and preferentially located in the proximity of the lip line. Structures of smaller size and presenting a more random orientation, on the other hand, can be observed in the instantaneous snapshots from the fractal jet. Therefore, the fractal orifice appears to break the large-scale, azimuthally-coherent structures, and redistributes the resulting energy among structures at smaller length scales. It can also be observed that the cross-plane velocity components are weaker in the fractal jet, consistent with the lower values of the rms of the radial velocity component in Fig.~\ref{fig:average-flow-vort}(e). Later in this paper, we will often be referring to Fig. \ref{fig:snapshots} to show how the analytical results that will be presented are supported by the qualitative observations from these instantaneous snapshots. 

\subsection{Analysis of the POD energy distribution}
\begin{figure}[ht]
    \includegraphics[width=0.9\textwidth]{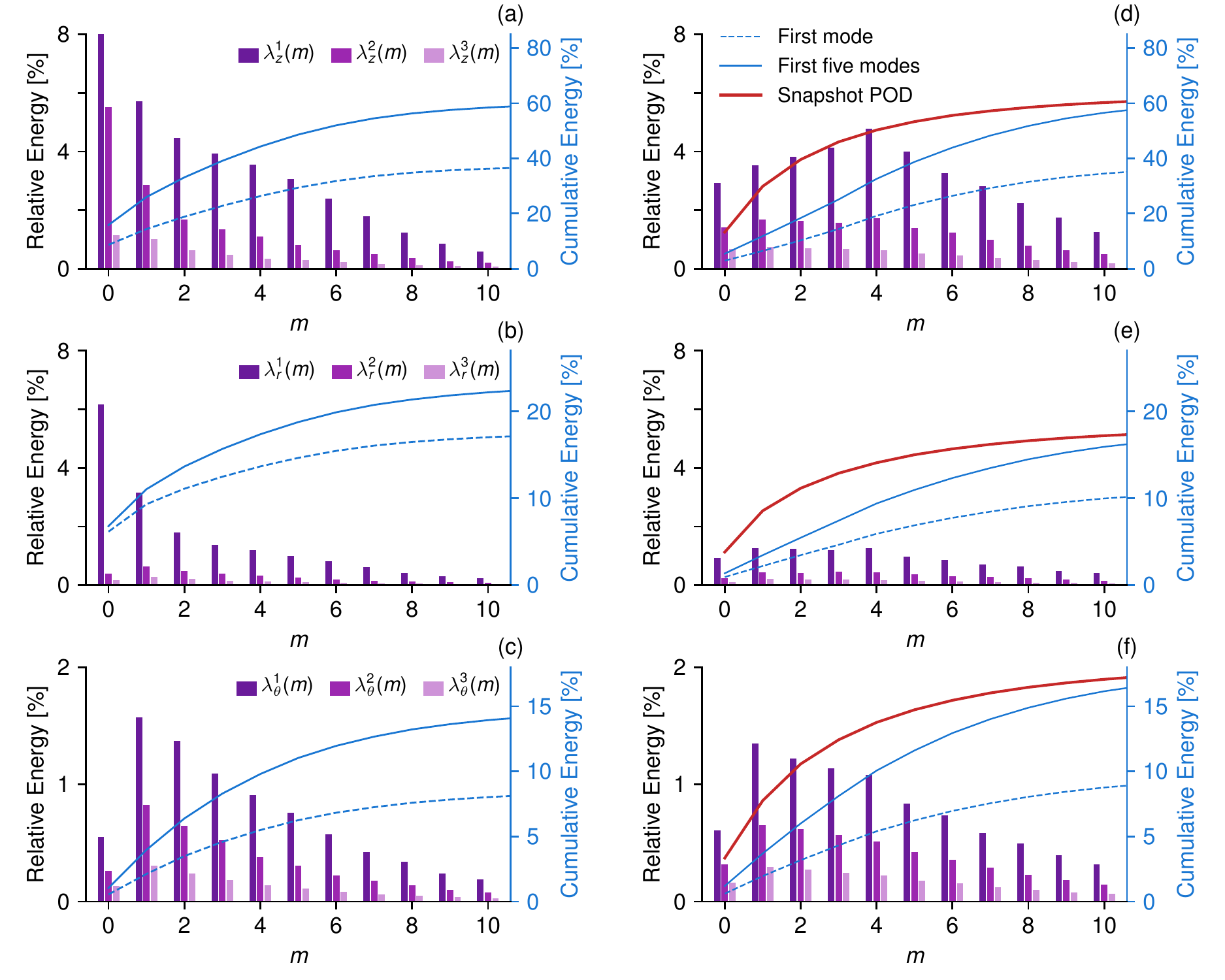}
    \caption{Relative energy distribution of the first three POD modes $i \in [1, 3]$ at the first eleven azimuthal modes, $m \in [0, 10]$, (a,c,e) for the round jet (left column) and (b,d,f) for the fractal jet (right column), at the downstream location of $z_0$. The POD analysis is made for the (a,b) streamwise (first row), (c,d) radial (second row), and (e,f) azimuthal (third row) components of the three-dimensional velocity vector field. The cumulative energy content of the first and first five POD modes are represented by a cyan dashed line and a cyan continuous line, respectively. For the fractal orifice, the cumulative energy of the first $10 (m + 1)$ snapshot POD modes is reported as a thicker red line, to quantify sub-optimality of the Fourier-POD analysis.}
    \label{fig:energy}
\end{figure}

To quantify how the turbulent kinetic energy at $z_0/D_{e} = 2$ is distributed across coherent structures at different wavenumbers, the Fourier-POD analysis described in section \ref{pod-method} is performed. A scalar implementation of this technique (see \textcite{Tinney2008}) is performed separately on each of the three velocity components. The relative fraction of turbulent kinetic energy captured by the first three POD modes for the first eleven azimuthal wavenumbers $m$ in each direction is presented in Fig.~
\ref{fig:energy} by vertical bars. The first, second, and third rows of the figure correspond to the streamwise, radial, and azimuthal velocity components, respectively, while the left and right columns show the results of the analysis for the round jet and for the fractal jet, respectively. The dashed cyan lines represent the cumulative energy distribution of the first POD modes as a function of the wavenumber $m$, whereas the continuous cyan lines represent the cumulative energy distribution for the first five POD modes. Note that quantities in this figure are relative respect to the overall energy across the three components. When examining Figs.~\ref{fig:energy}(a) and \ref{fig:energy}(d), the cumulative energy contained in the first ten POD modes and in the first eleven azimuthal wavenumbers of the streamwise velocity component is similar in the two orifices, i.e.~it is equal to approximately $60\%$. However, there are significant differences in the distribution of turbulent kinetic energy among the wavenumbers.
In the round jet, streamwise velocity fluctuations in the near field are dominated by the development of the Kelvin-Helmholtz instability in the axisymmetric vortex sheet, leading to the roll up of azimuthally coherent vortical rings in the transition region \citep{yule78}. This is observed in our results. The energy for the streamwise component is mostly concentrated at the wavenumber $m=0$, consistent with the results of \textcite{Jung2004}, see their figure 7. A rapid and monotonic decrease appears, though, for increasing $m$, which is slightly different from \textcite{Jung2004}, where the energy distribution of the first POD mode shows a local maximum at $m=6$. This subtle difference between our analysis and the study by \textcite{Jung2004} could be attributed to the lower Reynolds number of the present jet. The increase of the relative dominance of the first POD mode at the fundamental wavenumber for decreasing Reynolds numbers observed in figure 8 of \textcite{Jung2004} is consistent with this explanation. Another explanation for this difference is that the jet investigated by \textcite{Jung2004} is a nozzle jet while the jet examined here is an orifice jet. Therefore, the subtle decrease in the relative dominance of modes at $m=0$ could be a consequence of the \textit{vena contracta} effect associated with the orifice nozzle. In their experiments, \textcite{Iqbal2007} found that at $x/D=3$ the dominant wavenumber mode is $m=1$. \textcite{Iqbal2007} attributed the observed difference between their results and the results from \textcite{Jung2004} to the different initial conditions of the jet flows. 

The analysis of the jet issuing from the fractal orifice, Fig.~\ref{fig:energy}(d), reveals that the largest amount of streamwise fluctuation energy is contained at the fundamental wavenumber $m=4$. This is consistent with the fundamental geometric square pattern of the orifice, but is unrelated to the mean field of Fig.~\ref{fig:average-flow-vort}, since the Fourier-POD analysis is performed on the zero-mean velocity fluctuation fields. The dominance of the wavenumber mode associated with the orifice base-pattern is expected also for different orifice geometries, however less so when increasing the side number of the base pattern. Furthermore, in the fractal jet, the energy of the streamwise fluctuations is more scattered among the different azimuthal modes in the range $0 \le m \le 6$. As evident from the instantaneous snapshots in Fig.~\ref{fig:snapshots}, the physical mechanism at play is that the fractal geometry promotes the break-up of the near-field azimuthal coherence observed in the round jet. Hence, energy is injected at the fundamental wavenumber $m=4$ and is scattered across the wavenumber spectrum by the convective nonlinearity of the equations of motions. 

Analogous to the Fourier-POD analysis of the streamwise component, the analysis of the radial component reveals important differences between the round jet and the fractal jet, as can be appreciated from Figs.~\ref{fig:energy}(b) and \ref{fig:energy}(e). When looking at the cumulative energy distribution, it can be observed that the energy captured by the first five POD modes and by the first eleven azimuthal modes is larger in the round jet. Also, the cumulative energy from the first POD mode of each of the first eleven azimuthal wavenumbers (cyan dashed line) is almost double in the round jet than in the fractal jet. If we focus on the first POD mode of the round jet, we can observe that the zeroth and the first azimuthal wavenumbers, together, capture over $9 \%$ of the total energy. The observed dominance of the zeroth azimuthal wavenumber in the round jet is indicative of a low-dimensional behaviour of the radial component, which can be also appreciated in the instantaneous snapshots of Fig.~\ref{fig:snapshots}. Overall, the energy distribution among the modes obtained from the radial velocity component is consistent with previous experimental investigations with hot-wire anemometry from the literature \cite{Tinney2008, Iqbal2007, Jung2004}. In the fractal jet, however, the first POD mode accounts for a much smaller percentage of the total energy, independent of the wavenumber considered. The marginal importance of the radial POD modes in the fractal jet is a consequence of the reduced growth rate of the Kelvin-Helmholtz instabilities. As discussed in the recent stability analyses by \textcite{Lyu2019} and \textcite{Lajus2019}, the stronger is the deviation from the axisymmetry of the jet orifice, the lower is the growth rate of the near-field instabilities. For this reason, at the downstream position of $z_0/D_e = 2$, the energy of the radial velocity fluctuations is much larger in the round jet than in the fractal jet, as the rms in Figs.~\ref{fig:average-flow}(b) and \ref{fig:average-flow}(e) also show. 

The Fourier-POD analysis of the azimuthal component is presented in the third row of Fig.~\ref{fig:energy}. Here, the energy distribution appears almost insensitive to the orifice geometry. The wavenumber $m=1$ is dominant in both jets, with a monotonic decrease of the energy content of the modes for progressively higher wavenumbers. An analogous trend was found in previous Fourier-POD analyses of the azimuthal component from the literature \cite{Jung2004, Iqbal2007, Tinney2008}. The overall cumulative contribution of the first five POD modes and of the first eleven wavenumbers is only of approximately $15\%$ of the total energy. Therefore, among the three velocity components, the azimuthal component captures the lowest proportion of the turbulent kinetic energy. 

The energy distribution presented in Fig.~\ref{fig:energy} is obtained by applying a Fourier decomposition in the azimuthal direction followed by a POD analysis. This procedure has the advantage of enabling a straightforward comparison between the two jets, although it is sub-optimal for the fractal nozzle geometry, as explained in section \ref{pod-method}. Alternatively, the optimal modes could be obtained by a snapshot POD analysis. It is thus of interest to quantify possible changes in the cumulative energy distribution when this is computed from the modes obtained with the optimal snapshot POD analysis. The cumulative energy distribution of the first $10(m+1)$ modes obtained from the optimal snapshot POD analysis of the fractal orifice dataset is shown by the continuous red line in Figs.~\ref{fig:energy}(d, e, f). The Fourier-POD analysis, which is sub-optimal, leads to a cumulative distribution that only mildly underestimates the distribution obtained from the optimal snapshot POD analysis, supporting the relevance of the discussion on the Fourier-POD analysis. 

\begin{figure}
    \includegraphics[width=0.9\textwidth]{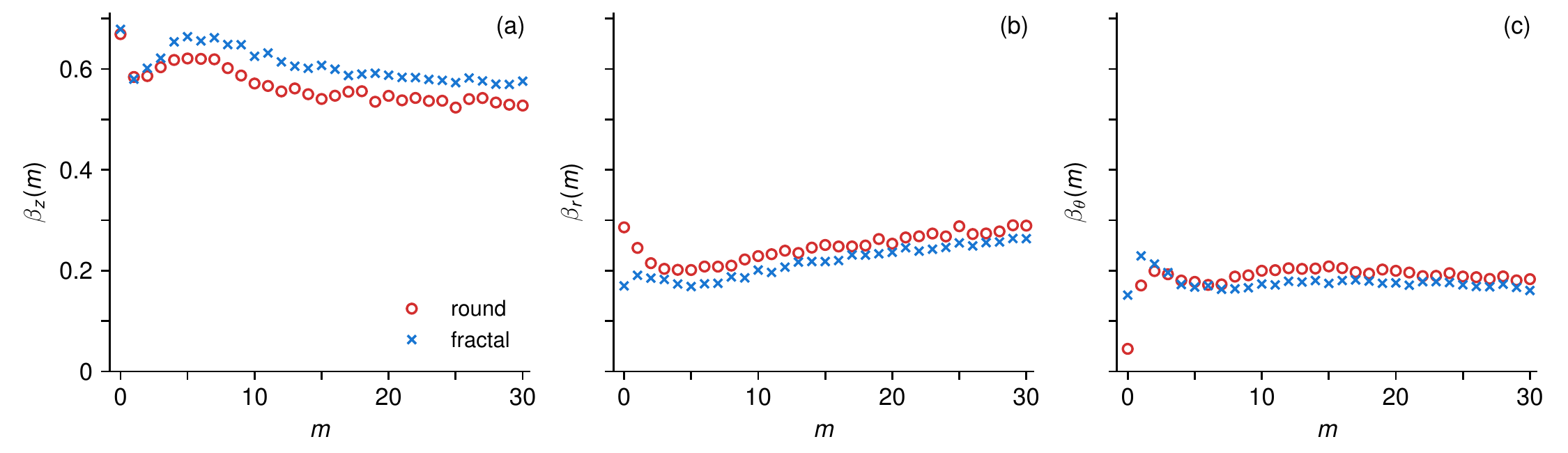}
    \caption{Relative energy contribution from the (a) streamwise, (b) radial, and (c) azimuthal velocity components to the energy contained within each of the first thirty wavenumbers. Red circles represent the round jet and cyan crosses the fractal jet. }
    \label{fig:energy-components-1}
\end{figure}

A more detailed quantification of the relative energy contributions of the POD modes of the different velocity components as a function of the azimuthal wavenumber is presented in Fig.~\ref{fig:energy-components-1}. The three panels show the ratios 
\begin{equation}
    \beta_{(\cdot)}(m) = \frac{\sum_{i}\lambda^i_{(\cdot)}(m)}{\sum_i \left [\lambda^i_z(m) + \lambda^i_r(m) + \lambda^i_\theta(m) \right ]},
\end{equation}
for the streamwise, radial and azimuthal components. As expected, and consistently with the results from Fig.~\ref{fig:energy}, the largest energy contribution to the total turbulent kinetic energy is obtained from the Fourier-POD modes of the streamwise velocity component. In Fig.~\ref{fig:energy-components-1}(a), the modes from the streamwise velocity component are responsible for approximately $70 \%$ of the energy contained in the wavenumber $m=0$, both in the round jet and in the fractal jet. However, this result should not be misinterpreted. The contribution of the wavenumber $m=0$ to the total turbulent kinetic energy is more than double in the round jet than it is in the fractal jet. This can be observed when comparing the cyan continuous lines in Figs.~\ref{fig:energy}(a) and \ref{fig:energy}(b). From Fig.~\ref{fig:energy-components-1}(b), as the wavenumber increases, the energy from the streamwise velocity component becomes gradually less important, whereas the radial component captures increasingly more energy. The most significant difference between the round jet and the fractal jet can be observed when assessing the proportion of energy taken by the radial and azimuthal components at low wavenumbers. In the round jet, the radial component captures around $30\%$ of the energy contained at wavenumber $m=0$, while the azimuthal component accounts for less than $5\%$ only. In the round jet, the proportion of energy from the azimuthal component is the lowest independent of the wavenumber. In the fractal jet, the same behavior holds, with the only exceptions of $m=1,2,3$.

\subsection{Analysis of the spatial structure of POD modes}
\begin{figure}[h!]
    \includegraphics[width=0.9\textwidth]{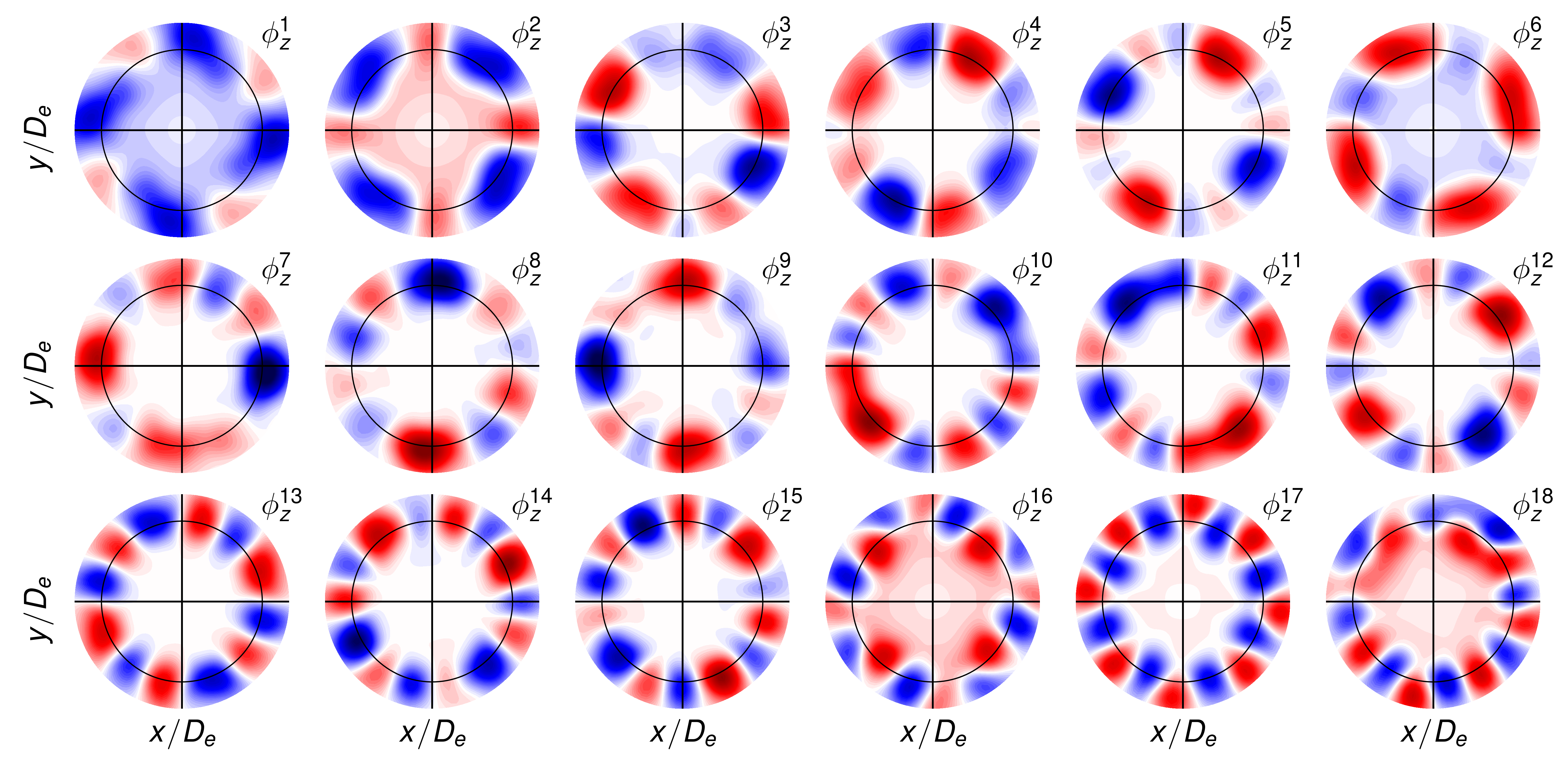}
    \caption{Streamwise component of the first eighteen snapshot POD modes in the jet with fractal orifice, on an arbitrary colour scale.}
    \label{fig:modes-fractal}
\end{figure}

The modes $\phi_z^i(r, \theta)$ of the first eighteen modes obtained from the snapshot POD of the streamwise component for the fractal orifice dataset are shown first in Fig.~\ref{fig:modes-fractal}. The structure of the two most energetic modes resembles the structure of the modes with $m=4$ obtained from the Fourier-POD analysis. This observation is consistent with Fig.~\ref{fig:energy}(b), where the modes at $m=4$ capture the largest amount of energy. The spectral content of the subsequent modes approximately follows the distribution reported in Fig.~\ref{fig:energy}(b). Thus, wavenumbers $m=2, 3$ and 5 all feature prominently in the first twelve modes. However, unlike the Fourier-POD analysis, the snapshot POD does not separate structures with different wavenumbers if these have similar energy and most of the POD modes of Fig.~\ref{fig:modes-fractal} have complex spectral characteristics.

\begin{figure}
    \includegraphics[width=0.86\textwidth]{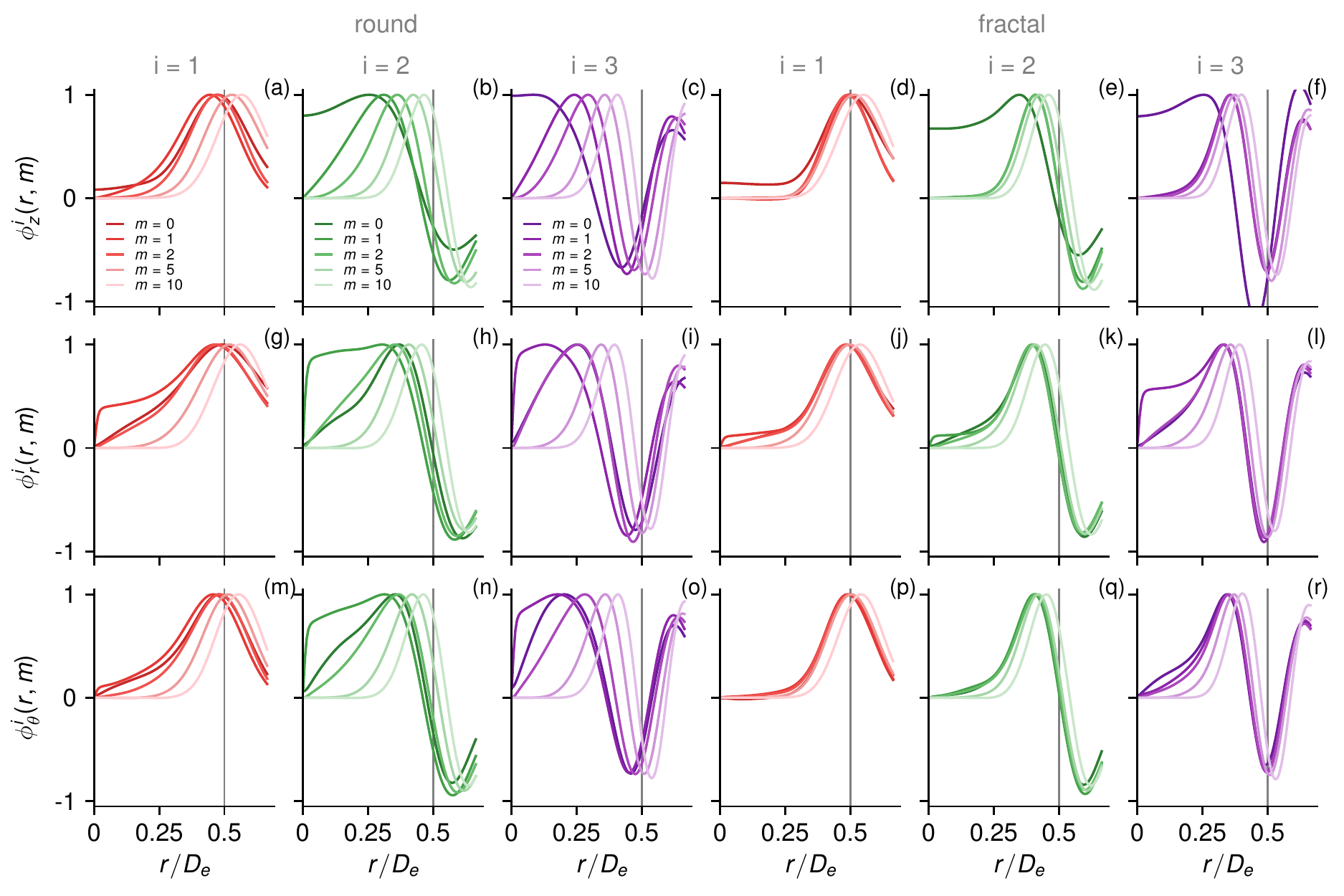}
    \caption{Radial profiles of the first three POD modes of the streamwise (first row), radial (second row), and azimuthal (third row) velocity components for azimuthal wavenumbers $m=0, 1, 2, 5, 10$, for the round jet (left columns) and for the fractal jet (right column). The grey vertical lines indicate the radial location $r/D_e = 0.5$. }
    \label{fig:radial-modes}
\end{figure}

To compare objectively the modal structures resulting from the two geometries, we examine the radial profiles of the first, second, and third Fourier-POD modes for azimuthal wavenumbers $m=0, 1, 2, 5, 10$, from the scalar analysis of the streamwise, radial, and azimuthal velocity components for the two jets. The profiles are normalised such that $\max_r\, \phi_{(\cdot)}^i(r, m) = 1$, and are presented in Fig.~\ref{fig:radial-modes}. In general, the $i$-th POD mode has $i$ extrema. For wavenumber $m=0$, the radial profiles of the first three POD modes tend to have a stronger support at low radial locations, approaching the centerline. This feature becomes stronger for increasing POD mode indices, and it is more prominent in the circular orifice than in the fractal orifice, as at $z_0/D_e=2$ the round jet presents a much larger energy content in proximity to the centerline (see Fig.~\ref{fig:average-flow}). The profiles of the first POD modes are similar to the profiles of the modes obtained with Spectral POD, Resolvent Analysis and transient growth analysis by \textcite{Nogueira2019}, reflecting the low-dimensional nature of the near-field jet dynamics.

\begin{figure}
    \includegraphics[width=0.88\textwidth]{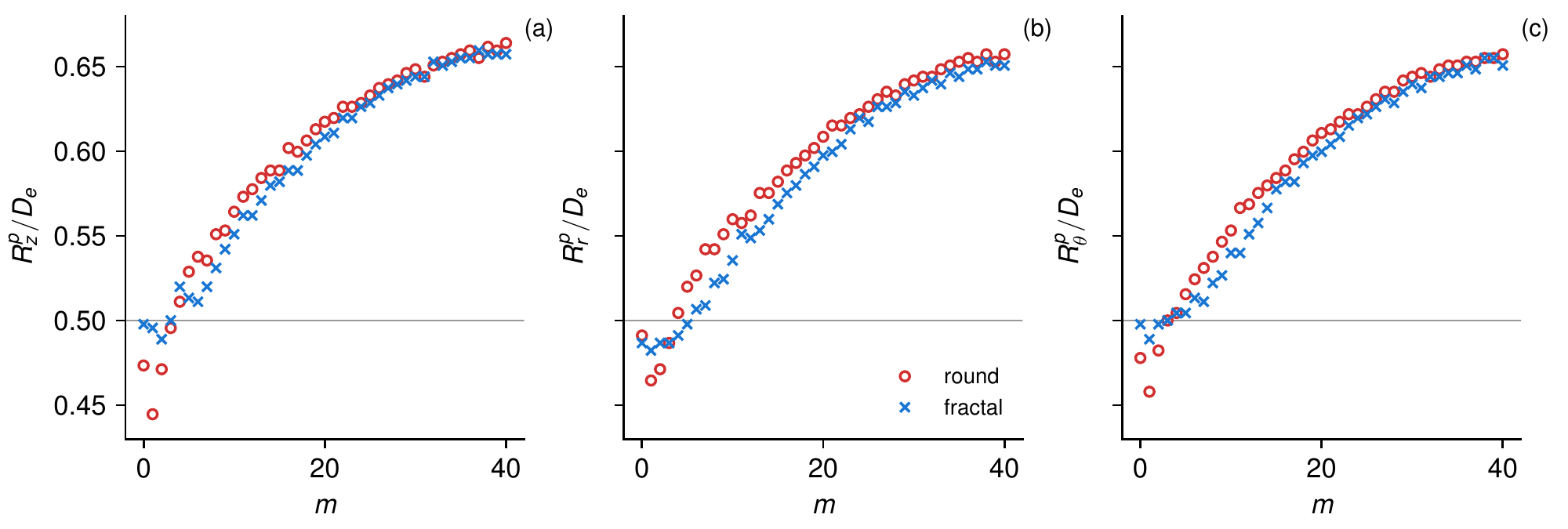}
    \caption{Normalised radial peak location of the radial profiles for the first POD mode of the (a) streamwise, $R^{p}_{z}/D_e$, (b) radial, $R^{p}_{r}/D_e$, and (c) azimuthal, $R^{p}_{\theta}/D_e$, velocity components as a function of the azimuthal wavenumber $m$, for the round (red circles) and fractal (cyan crosses) orifices. The grey horizontal lines indicate the radial location $r/D_{e} = 0.5$, i.e. the radial location of the edge of the round orifice. }
    \label{fig:collapse}
\end{figure}

The peaks of the radial profiles for the first, second and third POD modes tend to move outwards for growing azimuthal wavenumbers. However, this occurs more mildly for the fractal geometry, where the modal structures are more compactly localised around $r/D_e=0.5$. To quantify this behaviour, we denote by $R^{p}_{(\cdot)}$ the location of the first peak of $\phi_{(\cdot)}^i(r,m)$ from $r=0$. This quantity is shown in Fig.~\ref{fig:collapse} for the first POD mode as a function of the azimuthal wavenumber $m$, for the three velocity components and for the two orifice geometries. The peak location of the radial profiles exhibits analogous trends when computed from the three different velocity components. The first POD mode is radially localised around $R^{p}_{(\cdot)}/D_{e} \approx 0.5$, for the most energetic azimuthal wavenumbers, increasing to $R^{p}_{(\cdot)}/D_{e} \approx 0.65$ for high wavenumbers. The influence of the orifice geometry is mainly confined to the lowest wavenumbers, where the radial position of the peaks tends to be closer to the centerline. Circular and fractal orifices, however, present an analogous behaviour at $m>2$. This is not surprising, since the mean turbulent kinetic energy profiles in Figs.~\ref{fig:planar-profiles}(b) and (c) have comparable peak locations. An analogous radial displacement of the peak location was also observed in the Fourier-POD analysis of pipe flow reported by \textcite{Hellstrom2016} (their figure 2(a)).  As discussed by \textcite{Nogueira2019}, an important difference between jet and pipe flow, however, is that in pipe flow the radial expansion of the POD modes is limited by the wall, whereas in a jet the POD modes are not radially constrained and can expand beyond $r/D_{e}=0.5$. 

To further characterise the size of the POD structures, we introduce the radial half-size of the POD modes defined as $L^r_{(\cdot)} =( R^{p}_{(\cdot)} - R^{5}_{(\cdot)})/D_e$, with $R^{5}_{(\cdot)}$ being the radial location where $\phi_{(\cdot)}^i(R^{5}_{(\cdot)}, m) = 0.05 \phi_{(\cdot)}^i(R^{p}_{(\cdot)}, m)$. This parameter is more appropriate for an unbounded shear flow, as opposed to the parameter introduced by \textcite{Hellstrom2016} for pipe flow, where the radial length scale is defined as the distance from the wall to the peak location.
\begin{figure}
    \includegraphics[width=0.88\textwidth]{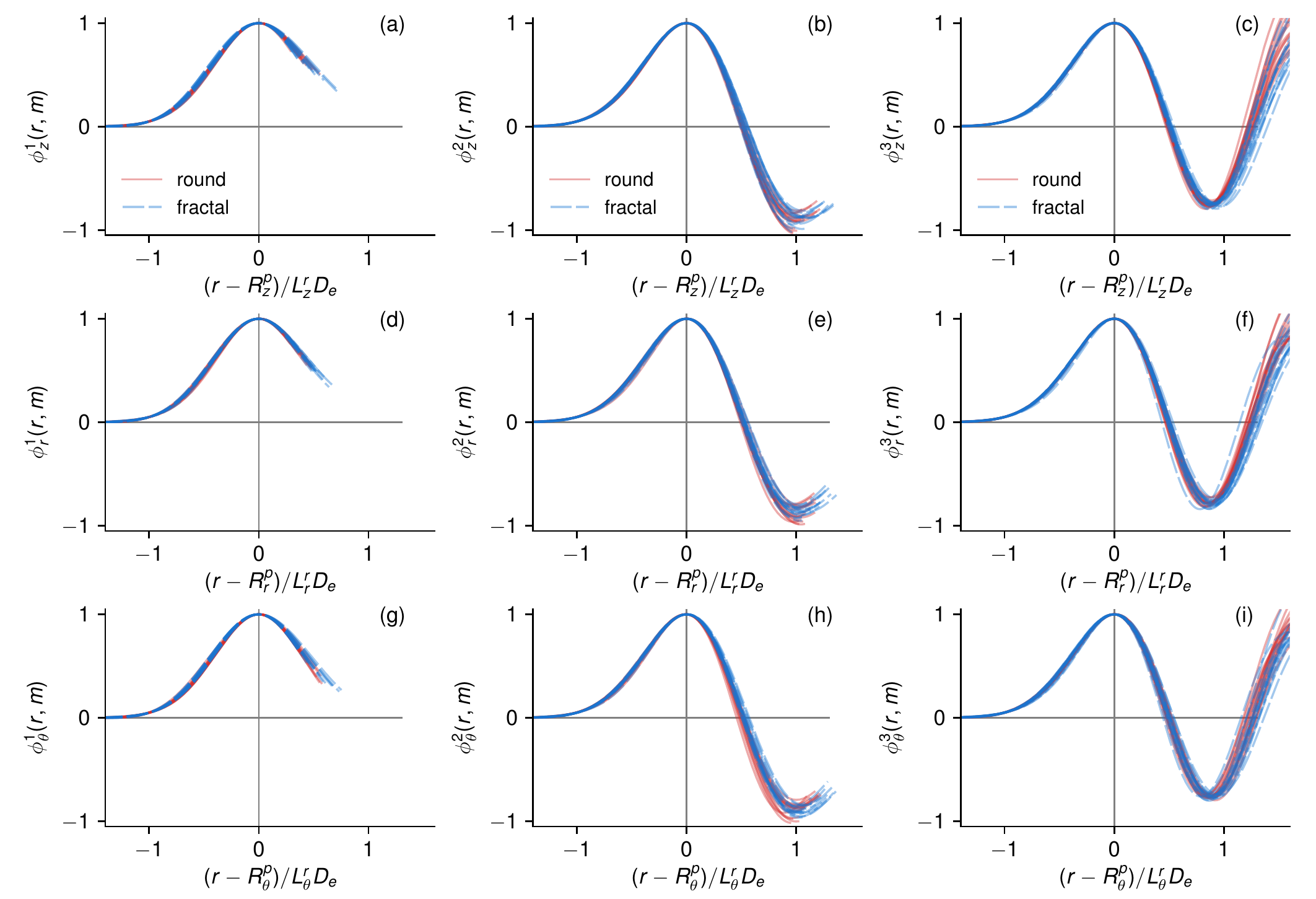}
    \caption{Radial profiles of the streamwise (first row), radial (second row), and azimuthal (third row) velocity components of the first, $i=1$, second, $i=2$, and third, $i=3$, POD modes for the azimuthal wavenumbers $m\in [2, 40]$, for the round jet (red continuous line) and for the fractal jet (cyan dashed line).}
    \label{fig:collapse-profiles}
\end{figure}
Modal structures identified by Fourier-POD analysis in pipe flow have been observed to collapse in the radial direction to a universal distribution when appropriately scaled with the radial length scale \cite{Hellstrom2016}. The same behaviour is observed in the present case.
% in the azimuthal direction, whereas the radial structures are extracted from the POD eigenvalue problem, and their properties should be examined in more detail. 
The collapse of the Fourier-POD profiles is shown in Fig.~\ref{fig:collapse-profiles}, for the first, second, and third POD modes for wavenumbers $m\in [2, 40]$. Profiles obtained from the POD analysis of the streamwise, radial and azimuthal velocity components are shown, for the round and fractal orifices. One salient observation is that for all azimuthal wavenumbers, the radial profiles collapse to a distribution that is surprisingly robust to the orifice geometry. It is argued that this is due to the similarity of the mean velocity profiles in Fig.~\ref{fig:planar-profiles}(a), since similar shear profiles produces fluctuations with similar spatial structure. 

%  I profili sono simili perche il profilo della velocita media sono simili. Mentre cambia la distribuzione di energy a causa della modulazione azimutale vedi cossu.

Another salient observation is that the profiles closely resemble those found in turbulent pipe flow by \textcite{Hellstrom2016}. Recently, analogies between the coherent structures in the near-field of a jet and in wall-bounded flows have been found by \textcite{Nogueira2019} and by \textcite{Samie2020}. In particular, following the self-similarity of POD structures reported by \textcite{Hellstrom2016}, \textcite{Nogueira2019} observed self-similarity in Spectral POD modes in the streamwise direction, where an azimuthal length scale is utilised to scale the streamwise development of the structures.
\begin{figure}
    \includegraphics[width=0.88\textwidth]{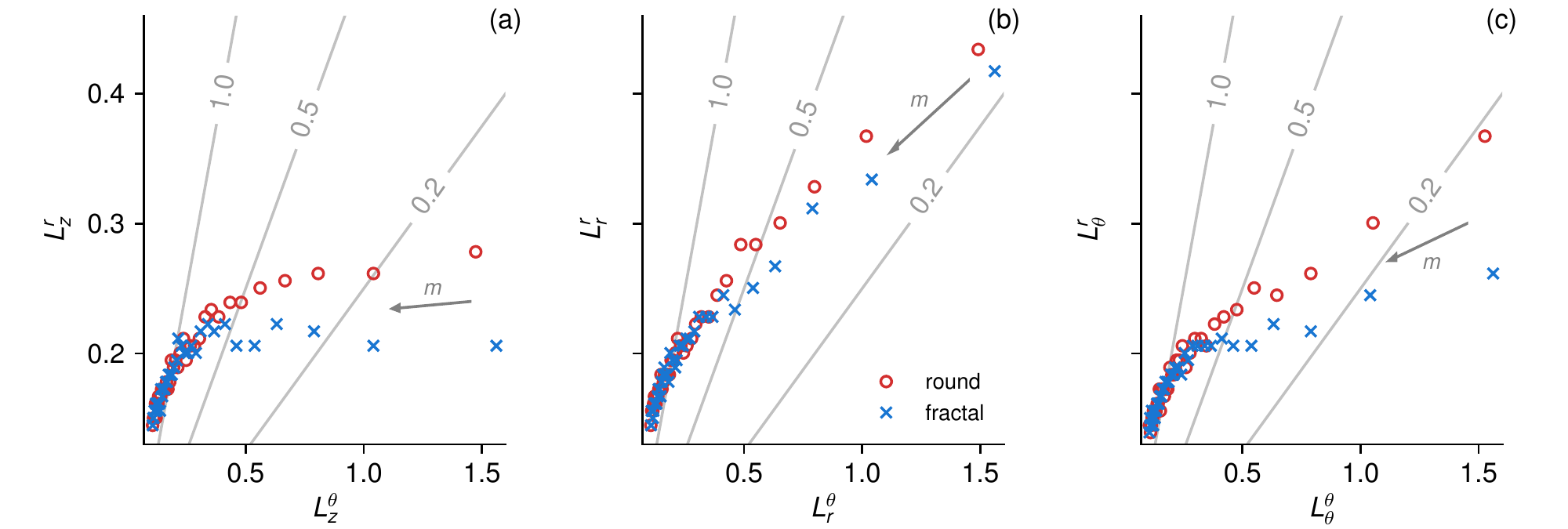}
    \caption{Radial length scale of first Fourier-POD modes as a function of their azimuthal length scale for the (a) streamwise, (b) radial, and (c) azimuthal velocity components. Data points for wavenumbers $m\in [1, 40]$ are shown, moving from right to left for growing wavenumbers, as shown by the arrows. The straight lines are used to estimate the aspect ratio of the eddies. }
    \label{fig:aspect}
\end{figure}
To examine the self-similarity of the first POD modes ($i=1$) in the present case, we introduce here the azimuthal length scale $L^{\theta}_{(\cdot)}=2\pi R^{p}_{(\cdot)}/(m D_e)$. The radial length scale of the POD structures as a function of the azimuthal length scale is presented in Fig.~\ref{fig:aspect}, for the (a) streamwise, (b) radial, and (c) azimuthal velocity components, for the first forty azimuthal wavenumbers $m\in[1, 40]$. A first remark on Fig.~\ref{fig:aspect} is that the POD structures associated to the first few, most energetic azimuthal wavenumbers of the fractal jet are more spatially compact, as the radial length scale $L^r_{(\cdot)}$ can be up to $\sim 25$\% smaller than in the canonical round orifice. The largest, more energetic structures have a radial support of about $0.4D-0.5D$, in agreement with the profile of mean turbulent kinetic energy of Figs.~\ref{fig:average-flow}(c, f). On the other hand, the radial support of the high-wave-number POD modes, $m \gtrsim 8$, does not seem to be heavily influenced by the orifice geometry, and it decays slowly for higher wavenumbers. These results suggest that the fractal orifice does not significantly affect the properties of the small scales, but it rather acts towards breaking the large-scale structures that would naturally appear in the round jet. In the figures, the straight lines identify structures with a constant aspect ratio  $L^{r}_{(\cdot)}/L^{\theta}_{(\cdot)}$. As can be observed, a univocal aspect ratio describing the eddies over the whole dynamic range of scales cannot be clearly defined, for none of the velocity components. This result makes the near-field jet different from pipe flow, where, according to \textcite{Hellstrom2016}, eddies are characterised by a constant aspect ratio of 0.2 across the wavenumber spectrum. 
This result is consistent with the spatial structure shown in the velocity snapshots in Fig.~\ref{fig:snapshots}, where eddies with small azimuthal length scale are not necessarily associated with smaller radial length scale. From a physical point of view, we argue that this behaviour arises from the absence of the wall-blocking effect present in pipe flow, which limits outward motions in the radial direction.

% The difference with the pipe structures appears to be mainly geometric, as $r$ is a spatial coordinate pointing outwards and not a wall-normal coordinate. 

\section{PROPER ORTHOGONAL DECOMPOSITION OF THE STREAMWISE VORTICITY AND VELOCITY}
\label{sec:coupling}

In the previous sections, Fourier-POD modes were identified for each velocity components separately, using a scalar POD implementation. 
% The spatial characteristics of the obtained modes were assessed, and the fraction of turbulent kinetic energy captured by each of the modes quantified. 
However, this approach does not capture possible correlations and couplings between the velocity components. Regarding this aspect, \textcite{Nogueira2019} showed that energetic streamwise velocity streaks in the high-shear region of the jet flow are strongly coupled to streamwise-elongated vortical structures by the lift-up mechanism \cite{BRANDT201480} as in wall-bounded shear flows \cite{Hamilton1995, Schoppa2002}, inducing strong radial motions. In this section, we examine the relationship between the streamwise velocity fluctuations and the structure of the streamwise vorticity field involved in the lift-up mechanism, to assess the effects of the orifice geometry.

\begin{figure}
    \includegraphics[width=0.96\textwidth]{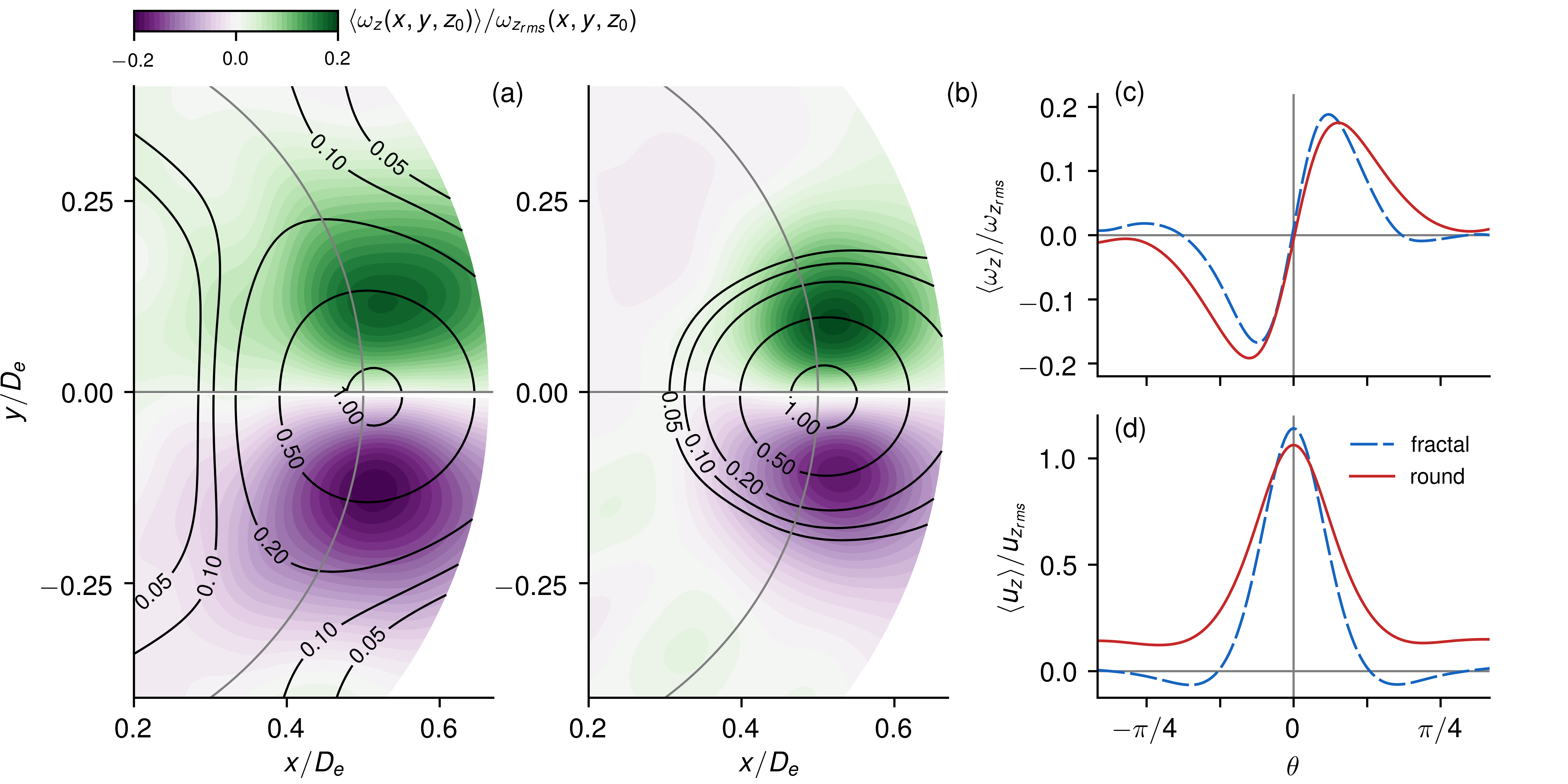}
    \caption{(a,b) Color maps of the normalised conditional average of the streamwise vorticity fluctuations, and labelled contours of the conditionally-averaged streamwise velocity, for both (a) the round jet and (b) the fractal jet (see text for details on the conditioning procedure); (c,d) radial profiles at constant $r/D_{e}=0.5$ as a function of the azimuthal angle $\theta$ of (c) the streamwise vorticity and (d) the streamwise velocity, from the conditional averages presented in the maps in (a,b). The grey lines in (a,b) trace the nominal lip line location.}
    \label{fig:conditional_average}
\end{figure}
 
We first show conditional averages of the streamwise vorticity and streamwise velocity fields. With the idea of characterising intense streamwise velocity fluctuation events, the condition for the average is for the value of the streamwise velocity fluctuation to be larger than a threshold value based on the rms of the streamwise velocity, $u_{z_{rms}}$, at the reference point $x/D_{e}=0.5$, $y/D_{e}=0.0$, $z/D_{e}=2.0$. This conditioning based on a threshold value can be expressed by the following equation:
\begin{equation}\label{thres}
 u^{\prime}_z > c \cdot u_{z_{rms}},
\end{equation}
where $c$ is a positive constant. The threshold value of the constant $c$ in equation \ref{thres} is chosen based on a sensitivity analysis, in which four different values of this constant were adopted, i.e. $c = 0.25, 0.5, 0.75, 1.0$. A similar condition can also be stated for negative fluctuations, but the probability distribution of the velocity fluctuations are not significantly skewed and lead to similar results. Although not presented here, the results of this analysis evidence that the obtained conditional averages are only moderately sensitive to the constant $c$, and that, as expected, larger values of $c$ lead to more intense events, with a smaller number of samples contributing to the averages. It was therefore chosen to set $c=0.5$ in equation (\ref{thres}), which represents a compromise between identifying  intense events, and retaining a sufficiently large statistical sample. In any case, the physics of the interactions remain unchanged for different constants $c$. A further improvement of the statistical significance is obtained by repeating the averaging for one hundred reference points distributed on the circumference at $r/D_{e}=0.5$, rotating the results back to $x/D_{e}=0.5$, $y/D_{e}=0.0$ and averaging along the azimuth. Conditional averages are denoted by angle brackets. Maps obtained from the described averaging procedure are presented in Figs.~\ref{fig:conditional_average}(a) and \ref{fig:conditional_average}(b) for the round and fractal jet, respectively. The colour map shows the conditionally-averaged vorticity field while the labelled contours show the conditionally-averaged streamwise velocity field. These variables are normalised by the maps of the rms of the respective quantities. The grey circle traces the nominal orifice lip line location. The averaging procedure unveils a pair of vorticity structures of opposite sign, flanking the point where the condition is defined, i.e.~in the proximity to the lip line. Therefore, positive streamwise velocity fluctuations are associated with positive fluctuations of the radial velocity component, i.e. a flow towards the jet periphery, that produces the mushroom-like structures observed in figure 9(a) of \textcite{Liepmann1992}. Similarly, although not presented here, negative fluctuations of the streamwise velocity are associated to a pair of high vorticity regions with opposite sign, with negative fluctuations of the radial velocity, i.e.~a flow towards the jet centerline. The described coupling between the sign of the streamwise velocity fluctuations at the lip line and the sign of the radial velocity fluctuations can also be observed from the instantaneous snapshots reported in Fig.~\ref{fig:snapshots}, in particular from the snapshot in Fig.~\ref{fig:snapshots}(c). Qualitatively similar results are obtained when the radial position of the point of condition is varied within the region of intense shear. The averaging captures a significant fraction, about 20\%, of the streamwise vorticity rms. Physically, high-speed fluid elements in the jet core are pushed outwards by radial motions produced by the streamwise vorticity features in regions of lower mean velocity, generating positive velocity fluctuations and vice versa. The observed structure of the conditionally-averaged streamwise velocity and vorticity fields supports the idea that a lift-up mechanism is active in turbulent jets \cite{Nogueira2019}.  In this scenario, the patches of coherent streamwise velocity derived from the conditional average can be regarded as the footprints of the streaks identified by \textcite{Nogueira2019} passing through the observation plane. If we compare the structures of streamwise vorticity obtained from the two orifice geometries, we can see that the vorticity structures from the round jet tend to have a larger extent in the azimuthal direction. This is more evident when examining Fig.~\ref{fig:conditional_average}(c), which shows radial profiles from the conditionally-averaged streamwise vorticity at constant $r/D_{e}=0.5$, as a function of the azimuthal angle $\theta$. It can be seen that the average vorticity structure within the fractal jet spans the range $-3 \pi/16 < \theta < 3 \pi/16$, while the structure within the round jet spans the range $-\pi/4 < \theta < \pi/4$. Fig.~\ref{fig:conditional_average}(d) shows the radial profiles from conditional averages of the streamwise velocity at constant $r/D_{e}=0.5$. For the round jet, the profile does not go to zero even for large azimuthal angles, consistent with the dominance of the azimuthal mode $m=0$ discussed in the previous section. Therefore, in the fractal jet the coupling between streamwise velocity and streamwise vorticity involves structures that are smaller in size, although of comparable strength. It should be stressed that the conditional averages presented up until here are obtained at the streamwise location $z/D_e=2.0$. If the same conditional averages were calculated at different streamwise locations, different structures of vorticity would be found, as the observed coupling between velocity and vorticity is strongly dependent on the streamwise position. However, this work focuses on examining the effects of the orifice geometry on the coupling between streamwise velocity and vorticity at a fixed downstream distance. Studying how this coupling varies with the streamwise position is left to future work.

The conditional averages presented above enable us to estimate the characteristic size of the vorticity structures sustaining the streamwise velocity fluctuations in both jets. With the aim of quantifying to what extent vorticity structures at different azimuthal scales contribute to the described mechanism, a vector implementation of the Fourier-POD analysis \cite{Tinney2008} is applied to composite snapshots of streamwise vorticity and streamwise velocity. The analysis identifies composite streamwise vorticity-velocity modes, and ranks them based on their importance. Note that, in what follows, we use the concept of modal energy, although this does not have the same dimensions of kinetic energy, as for the scalar POD implementation. Thus, the dominant vorticity-velocity are those that better capture intense and correlated fluctuations of the streamwise velocity and streamwise vorticity, on a wavenumber-by-wavenumber basis.
\begin{figure}
    \includegraphics[width=0.96\textwidth]{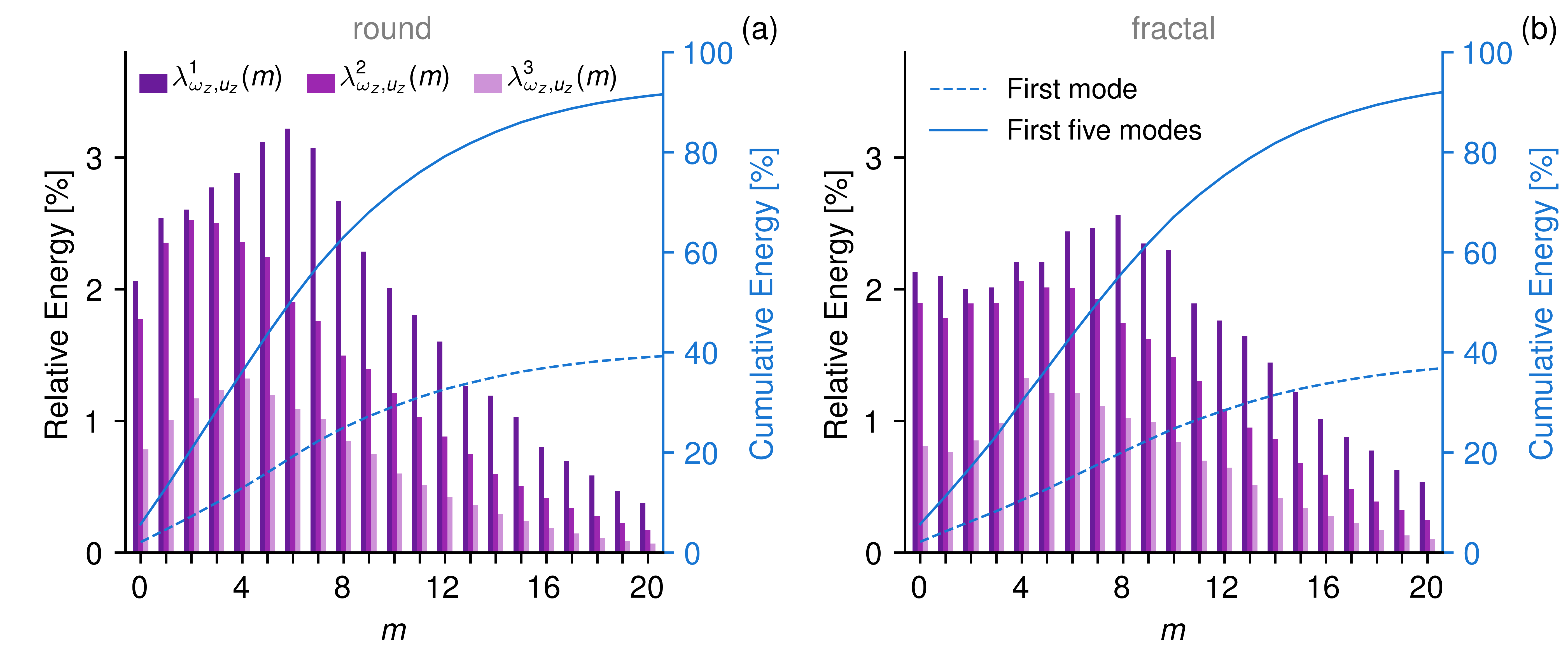}
    \caption{Relative energy distribution of the first three vector POD modes $i \in [1,3]$ at the first twenty-one azimuthal modes, $m \in [0,20]$, (a) for the round jet and (b) for the fractal jet. The Fourier-POD analysis is performed for the streamwise components of the vorticity and velocity fields. The cumulative energy content of the first and first five POD modes are represented by a cyan dashed line and continuous line, respectively.}
    \label{fig:vector-POD-energy-distribution}
\end{figure}
The relative energy distribution of the first three POD modes, $i \in [1,3]$, at the first twenty-one azimuthal modes, $m \in [0,20]$, is presented in Figs.~\ref{fig:vector-POD-energy-distribution}(a) and \ref{fig:vector-POD-energy-distribution}(b), respectively for the round jet and for the fractal jet. The dashed cyan lines represent the cumulative energy distribution of the first POD modes as a function of $m$, whereas the continuous cyan lines represent the cumulative energy distribution for the first five POD modes. As can be observed, the azimuthal wavenumber $m=6$ is dominant in the round jet, while the azimuthal wavenumber $m=8$ is the most important in the fractal jet. This result is consistent with the conditional averages of Fig.~\ref{fig:conditional_average}, where it was observed that smaller vorticity structures feature in the near field of the fractal jet. Fig.~\ref{fig:vector-POD-energy-distribution} also shows that the fractal orifice leads to a more scattered energy distribution across wavenumbers than the circular orifice, analogous to what found from the Fourier-POD analysis of the velocity, in the previous section. This property is also evident from cumulative energy distributions for the two jets. Although the first five POD modes and the first twenty-one azimuthal modes capture around $90\%$ of the total energy, modes at low wavenumbers (long length scales) are more important in the round jet than in the fractal jet, which leads to a steeper trend of the cumulative energy distribution in the former jet than in the latter one. 

\begin{figure}
    \includegraphics[width=0.9\textwidth]{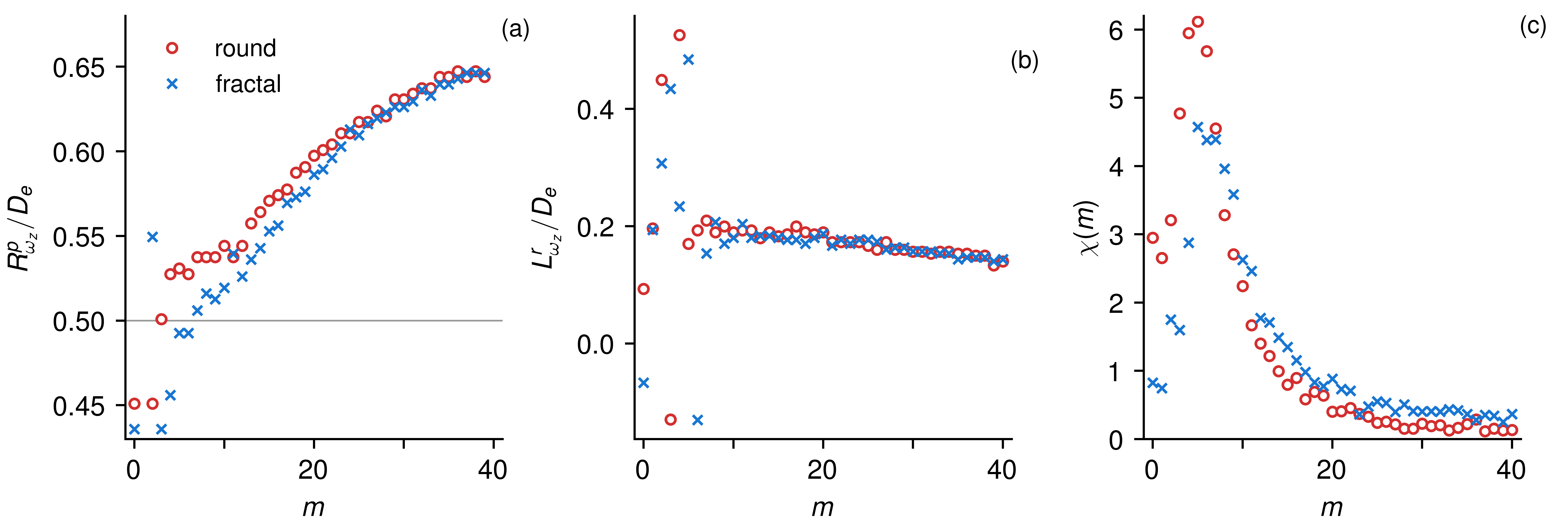}
    \caption{(a) Normalised radial peak location of the radial profiles of the streamwise vorticity component for the first vector POD mode; (b) normalised radial half-size of the vorticity component of the first POD modes as a function of the wavenumber $m$; (c) the quantity $\chi(m)$, defined in \eqref{eq:chi} as a function of $m$. }
    \label{fig:vector-POD-maxima}
\end{figure}

The normalised radial peak positions of the profiles of the vorticity component of the joint streamwise vorticity/velocity Fourier-POD modes are presented in Fig.~\ref{fig:vector-POD-maxima}(a) as a function of the azimuthal wavenumber. Analogous to what found in the analysis of the three velocity components, the peaks are located in the high-shear region and their locations move towards the periphery for increasing azimuthal wavenumber. Also, the peak positions for the fractal jet are nearer to the centerline when compared to the round jet. A reasonable explanation for this result is that the actual lip line of the fractal orifice reaches radial positions as close to the centerline as $r/D_{e} \approx 0.25$, introducing higher streamwise vorticity fluctuations in this area. The normalised radial length scale of the first POD vorticity structures $L^{r}_{\omega_{z}}/D_{e}$, defined similarly to the radial length scale of the velocity Fourier-POD modes, is presented in Fig.~\ref{fig:vector-POD-maxima}(b) as a function of the azimuthal wavenumber. Excluding the first four azimuthal wavenumbers, which define large-scale vorticity modes of little physical interest, the radial length scale of the POD structures decreases only moderately for increasing azimuthal wavenumbers, reaching values lower than $L^{r}_{\omega_{z}}/D_{e} \approx 0.2$. Although the dimension is lower than the radial length scale of the velocity POD modes discussed previously, the mild decrease of the radial length scale for $m>4$ indicates that a constant aspect ratio for the vorticity eddies cannot be determined. In fact, the azimuthal length scale $L^\theta_{\omega_z} = 2\pi R^p_{\omega_z} / m$ (not shown here) decreases much more rapidly with $m$ than $L^{r}_{\omega_{z}}$ and is thus not a suitable scale to characterise the POD profiles. As discussed, we argue that unlike in wall-bounded flows \cite{Hellstrom2016}, the lack of a solid wall originates a family of fluid structures that are free to occupy the entire radial dimension of the high-shear region, with little influence of the azimuthal wavenumber. This perspective is supported by the fact that this trend is essentially insensitive to the orifice geometry.

In Fig.~\ref{fig:vector-POD-maxima}(c), the ratio between the peaks of the velocity and vorticity components of the joint velocity-vorticity Fourier-POD modes, defined by the quantity
\begin{equation}\label{eq:chi}
 \displaystyle \chi(m) = 100 \times \frac{\max_r \phi^1_{u_z}(r, m)}{ \max_r \phi^1_{\omega_z}(r, m) },
\end{equation}
is presented as a function of the azimuthal wavenumber $m$ for both jets. This quantity is used here to quantify for the relative importance of streamwise velocity component, and as a proxy for the correlation between the two variables. It can be observed that $\chi(m)$ is largest for both orifices at wavenumbers in the range $5 \le m \le 8$. The fact that the velocity and vorticity correlation is larger at the wavenumbers where energy is mostly concentrated (see Fig.~ \ref{fig:vector-POD-energy-distribution}) corroborates the physical mechanisms described above, evidencing that a strong coupling between the streamwise vorticity and velocity is active in the near field.

\begin{figure}
    \includegraphics[width=0.7\textwidth]{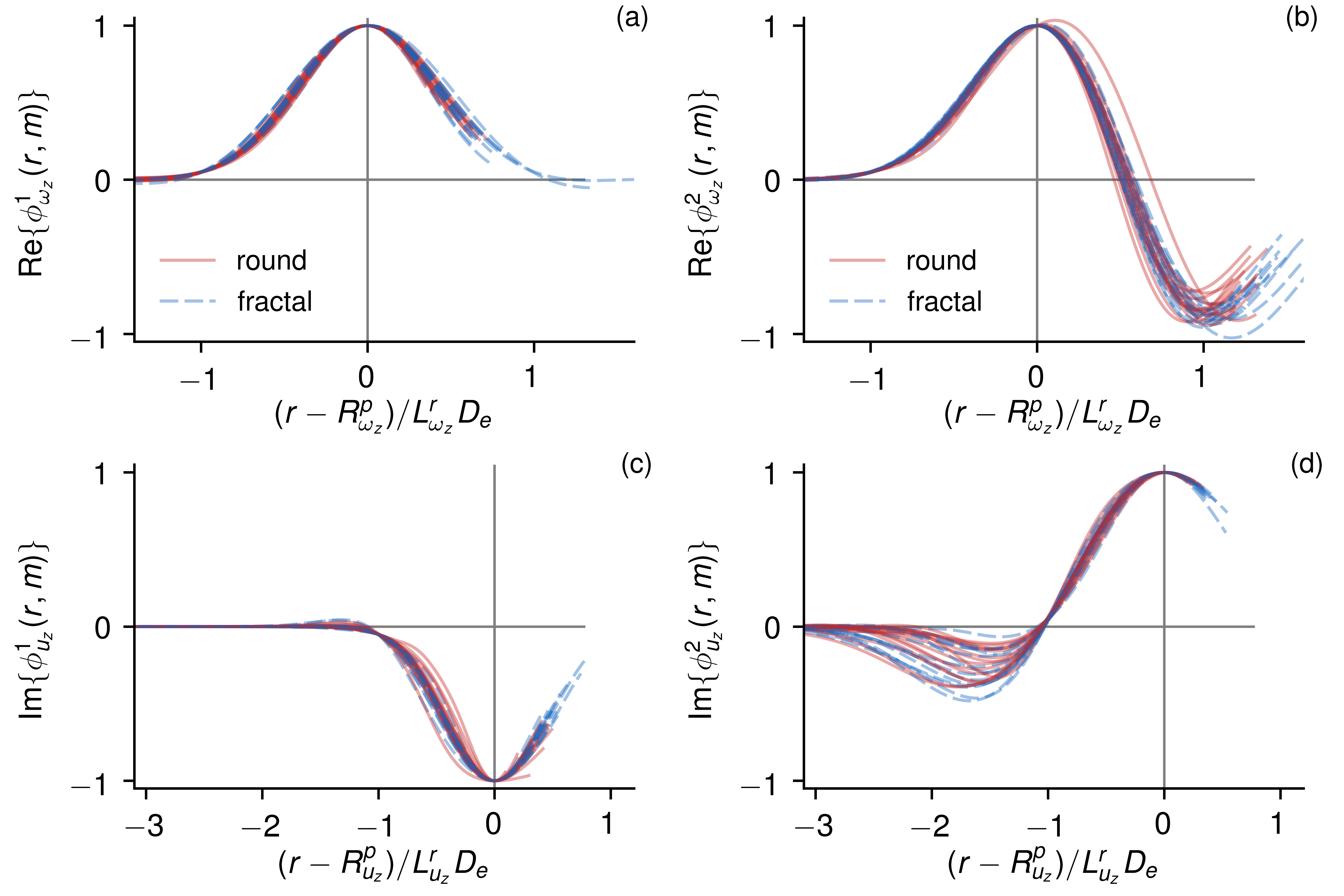}
    \caption{Radial profiles of the (a, b) streamwise vorticity and (c, d) streamwise velocity of the (a, c) first POD mode and (b, d) second POD mode, for the azimuthal wavenumbers $m\in [5, 40]$, for the round jet (red continuous line) and for the fractal jet (cyan dashed line). They grey vertical lines indicate the radial location $r/D_{e}=0.5$, i.e. the radial location of the edge of the round orifice. }
    \label{fig:vector-POD-scaled-profiles}
\end{figure}

Radial profiles of the vorticity and velocity component of the first two composite Fourier-POD modes are shown in Fig.~\ref{fig:vector-POD-scaled-profiles}, for wavenumbers $m\in[6, 20]$. The radial coordinate is scaled similarly to the profiles of Fig.~\ref{fig:collapse-profiles}, using the radial half-sizes $L^r_{\omega_z}$ and $L^r_{u_z}$. The real and imaginary parts of the Fourier-POD profiles are dominant for the vorticity and velocity components, respectively, and are thus reported in the figure. 
It can be observed that, similar to the velocity POD modes, the profiles collapse to a universal distribution that is independent of the orifice geometry. Profiles for smaller wavenumbers, corresponding to large-scale rotational motions with little dynamical relevance, deviate from such collapse. The relative arrangement of the vorticity and velocity components is more clearly understood by visualising the full structure of the Fourier-POD modes in the $x, y$ plane.
\begin{figure}
    \includegraphics[width=0.95\textwidth]{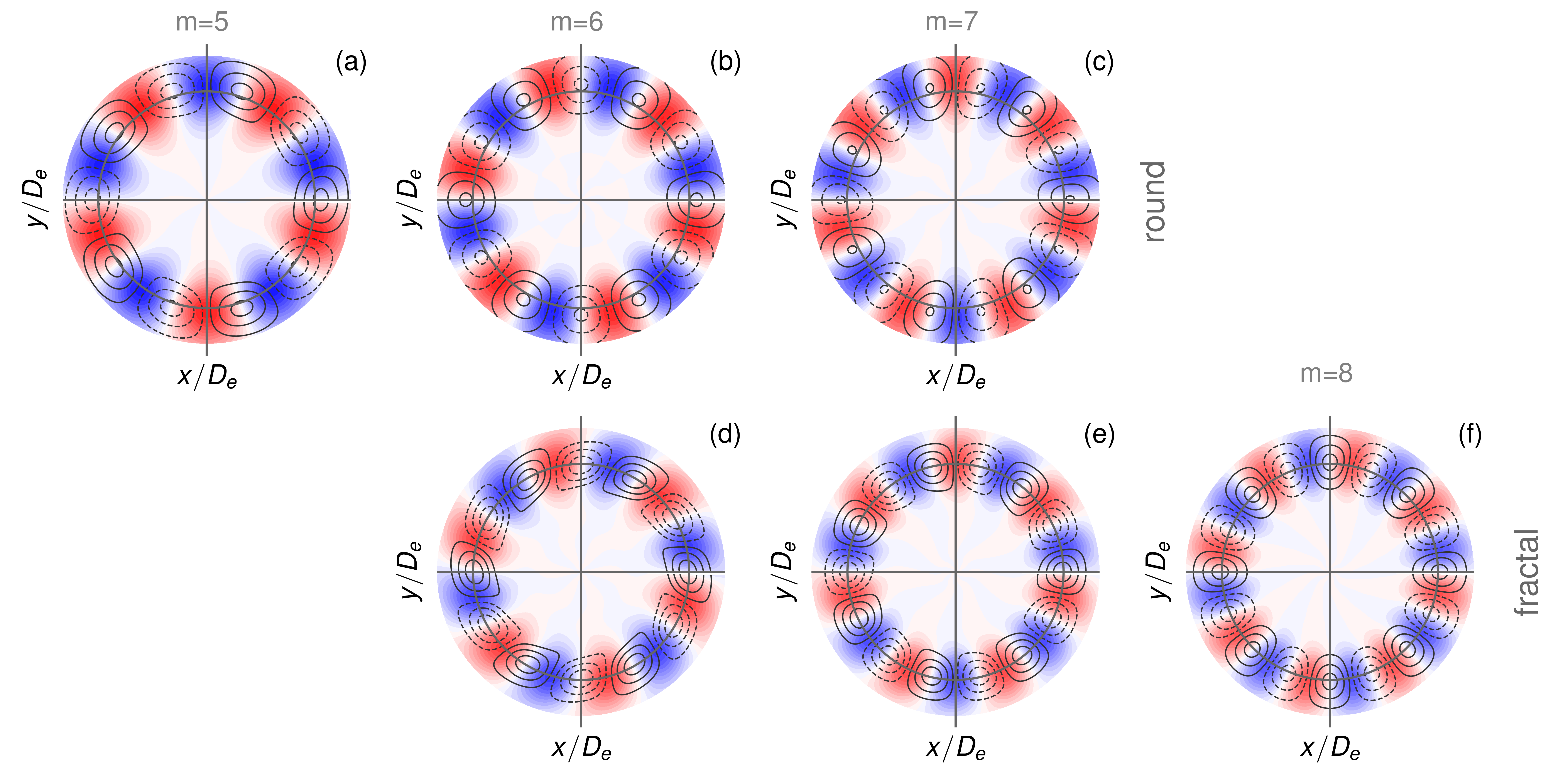}
    \caption{Streamwise velocity of the three most energetic Fourier-POD modes, on an arbitrary color scale, and contours of the streamwise vorticity, with continuous lines identifying positive values and dashed lines negative values, in the jet with (a, b, c) round orifice and with (d, e, f) fractal orifice.}
    \label{fig:vector-POD-mirini}
\end{figure}
The three most important Fourier-POD modes are presented in Figs.~\ref{fig:vector-POD-mirini}(a, b, c) and \ref{fig:vector-POD-mirini}(d, e, f), respectively for the round jet and for the fractal jet. Color maps of the streamwise velocity and contours of the streamwise vorticity are shown using arbitrary scales, since the modes are normalised. The same observations made on the relative spatial organisation of the conditionally-averaged vorticity and velocity fields apply to the Fourier-POD modes. Specifically, regions characterised by positive or negative fluctuations of the streamwise velocity are associated respectively with positive and negative fluctuations of the radial component, induced by pairs of vorticity structures. As observed, in the fractal jet, the azimuthal mode $m=8$ is dominant (Fig.~\ref{fig:vector-POD-mirini}(f)). From the maps of mean streamwise velocity and mean streamwise vorticity presented respectively in Fig.~\ref{fig:average-flow-vort}(d) and in Fig.~\ref{fig:average-flow-vort}(f), it can be observed that the mean flow characteristics dictate the relative importance of the vorticity-velocity Fourier-POD modes. In particular, this suggests that the spatial organization of both streaks and vortical structures is directly associated with the orifice geometry, consistent with the spectral POD analysis of \textcite{Rigas2019} on a jet from chevron nozzle. 

\section{Conclusions and future work}

In this work, a Fourier-POD analysis was performed to investigate the coherent structures in the near-field of a jet issuing from a non-circular orifice. More specifically, we considered an orifice with fractal geometry constructed from a base square pattern. These were compared with the structures in a jet issuing from a round orifice. The study considers three-dimensional velocity vector fields tomographic-PIV datasets obtained at a downstream distance from the orifice exit of two equivalent orifice diameters, which is used to characterize the role of the orifice geometry on the initial jet development.

From the analysis of the streamwise velocity component, the mode at wavenumber $m=0$, which captures the largest amount of turbulent kinetic energy in the jet with circular orifice, is not the dominant mode in the jet with fractal orifice. This is because the fractal geometry injects energy at the fundamental wavenumber $m=4$, thus breaking up the azimuthal coherence associated with the vortex rings typical of the round jet. As a result, while in the jet with circular orifice most of the energy is contained within these vortex rings, in the jet with fractal orifice the energy is distributed among the first seven azimuthal modes ($0 \le m \le 6$) more uniformly. Instantaneous snapshots of streamwise velocity confirm these findings, and show that structures within the fractal jet are smaller in size and lack of a preferential organisation. Consistent with this scenario, the energy in the radial component at the first wavenumber $m=0$, associated with the azimuthally-coherent radial motions from the Kelvin-Helmoltz vortex rings, is significantly lower in the fractal jet. In both cases, at the most energetic wavenumbers, the streamwise component was found to capture approximately $60\%$ of the total kinetic energy, followed by the radial component and by the azimuthal velocity components.

The radial distribution of the Fourier-POD modes from the two orifice geometries was also examined. It was found that the modal shapes at different wavenumbers collapse to a distribution that is independent of the orifice geometry, when scaled with a characteristic radial dimension. This finding is common to the Fourier-POD modes from each of the three velocity components. The most significant difference is in the radial support of the modes of the jet with fractal orifice being smaller at low wavenumbers. This evidences that the most energetic structures tend to be smaller in size in the jet with fractal orifice. Regardless, the orifice geometry does not appear to significantly affect the shape of the first POD modes, but mostly the energy distribution. The collapse of the radial profiles was also observed in recent experiments in turbulent pipe flow \cite{Hellstrom2016}. However, an important difference with these observations is that, in the present case, the ratio between the azimuthal and radial length scales of the Fourier-POD structures varies with the wavenumber $m$, while it is constant in pipe flow and approximately equal to $0.2$. This result was explained by the fact that in jet flow the lack of a wall-blocking effect does not significantly constrain the radial extent of turbulent motions. In this respect, while \textcite{Nogueira2019} found that the SPOD structures are self-similar in the streamwise/azimuthal directions, the aspect ratio in the radial/azimuthal directions was not examined. The present investigation of the radial/azimuthal aspect ratio contributes therefore to complete the chararacterization of coherent structures in turbulent jets. 

Streaky structures have been recently found in the near field of a jet, resulting from a lift-up mechanism analogous to turbulent wall-bounded flows (\textcite{Nogueira2019}). The structures of streamwise vorticity leading to streamwise velocity fluctuations were examined in relation to the orifice geometry. To this aim, \textit{i.}) an averaging of the streamwise vorticity conditioned on the intense streamwise velocity fluctuations at points on the nominal lip line location, and \textit{ii.}) a joint Fourier-POD analysis of streamwise velocity and streamwise vorticity were performed. Conditional averaging showed that intense positive fluctuations of the streamwise velocity are associated with pairs of streamwise vorticity structures of opposite sign, flanking the point of conditioning. Consistent with streak/roll dynamics in wall-bounded shear flows, the combined activity of this vortex pair induces positive fluctuations of the radial velocity, i.e. a flow towards the jet periphery. Alternatively, negative fluctuations of the streamwise velocity are associated to negative fluctuations of the radial velocity, i.e. a flow towards the jet centerline. The flow pattern obtained by this averaging procedure is analogous for both the orifice geometries, although the vorticity structure from the round jet is $30\%$ larger than the vorticity structure from the fractal jet. From the joint Fourier-POD analysis, the fractal orifice promotes the involvement of structures over a wider range of length scales in the mechanism leading to the streaks formation. Lower wavenumber modes contribute in larger part to the overall fluctuation budget in the round jet compared with the fractal jet. This is consistent with the larger size of the vorticity patterns from conditional averaging being larger in the jet from the circular orifice. These aspects have been examined only at two diameters from the orifice, and further investigations on the axial development of the coupling between streamwise velocity and streamwise vorticity are warranted. However, our expectation is that the coupling becomes milder as the jet develops axially, due to the progressively decreasing mean shear at larger streamwise distances.

Recent observations both from experiments (\textcite{Jordan2018}) and from numerical simulations (\textcite{Towne2017}) underline the importance of the Kelvin-Helmholtz instabilities in the generation of tonal noise from jet-flap interactions. According to these studies, the wavepackets associated with the Kelvin-Helmholtz instabilities resonate with upstream-travelling trapped acoustic modes located in the potential core, thus producing far-field tonal noise. In relation to these studies, the attenuation of the Kelvin-Helmholtz instabilities by the non-circular orifice could mitigate the effects of this resonance, and ultimately reduce the emissions of tonal noise originating from the interactions between a round jet and the edge of a flap. Future studies of jet-flap interaction noise with a jet nozzle of non-circular geometry could shed further light on these aspects. 

\bibliography{JET} 

\end{document}